\documentclass{aa}  
\bibpunct{(}{)}{;}{a}{}{,}
\usepackage{graphicx}
\usepackage[switch]{lineno} 
\usepackage{hyperref}
\usepackage{threeparttable}
\usepackage{txfonts}
\newcommand{\maps}{\textit{Meteoritics \& Planetary Science}}
\newcommand{\psj}{\textit{The Planetary Science Journal}}
\begin{document} 


   \title{Mid-infrared absorption spectra and mass absorption coefficients for 23 chondrites}

   \subtitle{Dependence on composition and grain size}

   \titlerunning{Mid-infrared spectra of 23 meteorites}

   \author{Grace A. Batalla-Falcon\inst{1,2},
          Lucas A. Cieza\inst{1,2,3},
          Roberto Lavin\inst{1,2,4},
          Millarca Valenzuela\inst{5,6},
          Andreas Morlok\inst{7},
          Prachi Chavan\inst{1,3},
          Cecilia Farias\inst{2},
          Giovanni Leone\inst{8}
          \and
          Daniel Moncada\inst{9}
          }
   \authorrunning{Batalla-Falcon et al.}
   
   \institute{Instituto de Estudios Astrofísicos, Facultad de Ingeniería y Ciencias, Universidad Diego Portales, Santiago, Chile\\
              \email{grace.batalla@mail.udp.cl}
         \and
             Cosmic Dust Laboratory, Universidad Diego Portales, Santiago, Chile
        \and
             Millennium Nucleus on Young Exoplanets and their Moons (YEMS), Santiago, Chile
        \and
             Instituto de Ciencias Básicas, Facultad de Ingenería y Ciencias, Universidad Diego Portales, Santiago, Chile
        \and
             Departamento de Ciencias Geológicas, Universidad Católica del Norte, Antofagasta, Chile
        \and
             Centro de Astrofísica y Tecnologías Asociadas (CATA), Macul, Chile
        \and
             Institut für Planetologie, Westfälische Wilhelms-Universität Universität Münster, Germany
        \and
             Instituto de Investigacion en Astronomía y Ciencias Planetarias, Universidad de Atacama, Copiapó, Chile
        \and
             Departamento de Geología, Facultad de Ciencias Físicas y Matemáticas, Universidad de Chile, Santiago, Chile
             }



  \abstract
   {}
   {We present mid-infrared transmission spectra from 2 to 23 \textmu m of the 23 Atacama Desert chondrites of different types (carbonaceous Ornans and ordinary of H, L, and LL groups) as well as of some pure minerals (olivine and diopside). We focus on the characteristics of silicate at 10 \textmu m and 20 \textmu m, analyzing the influence of composition and grain size on peak strengths and spectral shapes. We present the first results of the Cosmic Dust Laboratory, a dedicated facility at the Universidad Diego Portales equipped with a VERTEX 80v vacuum Fourier transform infrared spectrometer.}
   {Through milling and sieving samples, we obtained different ranges of particle sizes to study the effect of grain size on the intensity and shape of the spectrum.}
   {The resulting spectral library can be compared with astronomical data of protoplanetary disks, debris disks, and even white dwarf disks obtained with instruments such as MIRI on board the \textit{James Webb} Space Telescope and MATISSE on the Very Large Telescope Interferometer. We also present mass absorption coefficient values, which can be used for radiative transfer modeling of astronomical observations. This study aims to improve dust opacities for astronomical applications, with a focus on circumstellar disks.}
   {}

   \keywords{dust --
                meteorites --
                spectroscopy --
                grain size --
                protoplanetary disk --
                protoplanetary dust 
               }

   \maketitle
%
\section{Introduction}
    According to the nebular theory postulated centuries ago by Immanuel Kant, the planets in our Solar System formed in a rotating disk of gas and dust \citep{1755anth.book.....K}. This idea was found to be essentially correct. In fact, protoplanetary disks - analogous to the solar nebula that produced our Solar System - are now routinely studied using the world's most powerful telescopes, including the Atacama Large Millimeter/Submillimeter Array (ALMA), the Very Large Telescopes (VLT), and the \textit{James Webb} Space Telescope (JWST). 
Furthermore, recent exoplanet surveys from the ground and from space have found that extrasolar planets are ubiquitous in the Milky Way \citep{2015ARA&A..53..409W}. In particular, the extremely successful Kepler mission has demonstrated that rocky planets such as Earth and mini-Neptunes are much more common than giant planets such as Jupiter or Saturn \citep{2011ApJ...736...19B, 2023ASPC..534..839L}. However, planet formation remains an unsolved problem, as it is a very complex process involving the growth of submicron-sized dust particles into large differentiated bodies that are thousands of kilometers across.

    Some of the first steps in planet formation are particularly problematic \citep{2010AREPS..38..493C} since it is still not understood how small dust aggregates (pulled together by electrostatic sticking) can grow into kilometer-sized bodies massive enough to allow for oligarchic growth (at which point gravity plays a crucial role). Several planet formation barriers exist between the electrostatic and oligarch growth regimes, including the fragmentation barrier and the radial drift barrier. The former refers to the tendency of rocks to fragment (instead of growing) when they collide, and the latter refers to the very rapid inward migration that is expected for dust particles when they reach a certain size (a few millimeters) in a gas-rich disk \citep{2008A&A...480..859B}. 

    Protoplanetary disks are the current equivalents to the early Solar System \citep{2011ARA&A..49...67W}, making them a particular subject of focus in modern astrophysics \citep{2020ARA&A..58..483A}.  
    Although circumstellar dust typically represents only 1$\%$ of disk material by mass \citep{2007EJMin..19..771J, 2010ConPh..51..381W}, it is the main observable because it dominates the opacity at infrared and (sub)millimeter wavelengths. Astronomical observations trace dust grains with sizes similar to the observed wavelengths (e.g., a few \textmu m for JWST and $\sim$1 mm in the case of ALMA).
    Interpreting astronomical observations thus requires a detailed understanding of protoplanetary dust.
    In particular, the mid-infrared (MIR) spectrum of this dust depends on different parameters, such as chemical composition, shape, roughness, temperature, and particle size \citep{1963EcGeo..58..274L,1986IJRS....7.1879S}.

    Experimental results have clearly shown that chemical variations in mineral phases, such as olivine, pyroxene, melilite, or plagioclase, which are the main components of circumstellar dust, have different absorption spectra in the MIR \citep{2000A&A...363.1115K, 2017ApJ...845..115K,  2002A&A...391..267C}. 
    The spectral features of dust in the MIR spectra of protoplanetary disks are typically analyzed using dust models that incorporate spectral libraries of pure minerals, such as amorphous silicates, forsterite, and enstatite. \citep{2023ApJ...945L...7K, 2024ApJ...963..158P, 2024A&A...691A.148J}. 
    In addition, using analytical approaches, most studies extrapolate the opacity libraries in grain size and/or wavelength.  

    Rather than studying artificial mixtures of terrestrially formed minerals (as in the works cited in the previous paragraph), planetary and extraterrestrial materials are likely more physically and compositionally “realistic” analogs to protoplanetary disks. More specifically, their current composition, mineralogy (e.g., silicates), and texture represent a combination of (i) the material they accreted in the disk and (ii) the diverse postaccretional processes that further occurred during the early Solar System and protoplanetary phase (e.g., thermal metamorphism, differentiation, impacts). Creating artificial laboratory mixtures that recreate all of these processes or trying to reproduce them with models are complex and challenging tasks. Moreover, certain physical properties of extraterrestrial materials may differ from those of models of bare mineral mixtures. For instance, interplanetary dust particles (IDPs) have been found to have organic coatings that increase the stickiness coefficient of the material compared to bare minerals.
    In fact, parameters such as the stickiness coefficient, which plays a significant role in the aggregation of the first dust particles in the solar nebula \citep{2013EP&S...65.1159F} are often excluded in works using artificial mineral mixtures.

    Concerning more realistic samples, it is possible to have access to different extraterrestrial particles in order to perform measurements in the laboratory. Interplanetary dust particles can be found in asteroids and comets, and they can even fall directly on Earth. Organic coatings have been found in chondritic porous IDPs collected from the Earth’s stratosphere by NASA high-altitude research aircraft. These coatings increase the sticking coefficient compared to bare mineral grains and serve as a significant aid in the aggregation of the first dust particles in the solar nebula \citep{2013EP&S...65.1159F}.
%
    Moreover, samples returned by the most recent space missions, such as the Hayabusa mission \citep{2011Sci...333.1113N}, have been beneficial to different laboratory measurements. 
    Laboratory spectral measurements have been conducted on a variety of extraterrestrial materials, ranging from IDPs \citep{1985ApJ...291..838S, 2020PSJ.....1...62M}, particles from sample return missions \citealt{2011Sci...333.1113N, 2015M&PS...50.1562B}, and meteorites \citep{1991Icar...92..280S, 2010GeCoA..74.4881B, 2014Icar..229..263B}. However, meteorites are the most available sources of extraterrestrial materials on Earth, particularly stony meteorites. Due to their silicate groundmass, stony meteorites are a compatible option for the study of  protoplanetary disks in the MIR, as the latter predominantly display broad silicate emissions at MIR wavelengths  \citep{2008M&PS...43.1147M, 2012Icar..219...48M, 2014Icar..231..338M, 2014Icar..229..263B}. By extension, powders of stony meteorites should be good analogs to dust in protoplanetary disks.

    Meteorite classification is based on chemical composition and petrogenesis, grade of weathering, and grade of shock metamorphism. The most relevant parameters used to classify meteorites are petrographic characteristics (texture, mineralogy, and mineral composition), whole-rock chemistry, and oxygen isotopic (O-isotopic) composition \citep{2006mess.book...19W}. 
Based on their chemical composition and petrogenesis, stony meteorites can be separated into three main categories: undifferentiated meteorites, known as "chondrites", which can be either unequilibrated or equilibrated meteorites; differentiated meteorites, which include "achondrites" (a subset of stony meteorites) as well as iron meteorites and stony-iron meteorites; and an intermediate stage referred to as "primitive achondrites" \citep{2006mess.book...19W, 2014mcp..book...65S, 2014mcp..book....1K}. This classification reflects the range of processes that meteorites can experience, from minimal alteration to extensive differentiation and melting. Chondrites correspond to meteorites that usually contain small spherical mineral agglomerates called chondrules that can vary in size from 0.02 mm to 10 mm from one class of meteorite to another \citep{2006mess.book...19W}. Chondrites have similar solar elemental abundances (with depletion of volatile elements) and are derived from asteroids that do not experience differentiation. Achondrites correspond to igneous rocks or breccias of igneous fragments that have suffered differentiation processes: melting, partial melting, or melt residues \citep{2020M&PS...55..857T}. Primitive achondrites correspond to meteorites that present an achondritic texture (igneous or recrystallized); still, their chemical composition is more similar to the precursor of the chondrite and is considered closer to their primitive parent body than achondrites \citep{2006mess.book...19W}. 

    Chondrites are subdivided into three main classes: carbonaceous chondrites (CCs), ordinary chondrites (OCs), and enstatite chondrites (E). In addition, there are two less common classes known as kakangari chondrites (K) and rumuruti chondrites \citep[R;][]{1996GeCoA..60.4253W, 2011ChEG...71..101B, 2014mcp..book....1K}. 
    A meteorite class consists of groups that share a similar primary bulk composition, O-isotopic properties, and other chemical characteristics, such as abundances of refractory lithophile elements \citep{2013GeCoA.108...45W}. In addition to these chemical properties, a meteorite class also tends to exhibit overall petrographic similarities, including chondrule size, abundances of refractory inclusions, and matrix abundances \citep{2004mete.book.....H, 2006mess.book...19W, 2014mcp..book....1K}. 
    It is important to note that OCs are by far the most abundant chondrite class among all discovered meteorites. Although carbonaceous asteroids are more abundant in the Solar System \citep{1982Sci...216.1405G, 2013Icar..226..723D, 2020SSRv..216...55K}, there is a noticeable bias toward the collection of OCs over CCs. This bias is primarily due to the higher survivability of OCs during atmospheric entry and their greater resistance to weathering on the ground \citep{1998SSRv...84..327C, 2018ChEG...78..269F, 2018M&PS...53.2181R}. Despite CCs being the dominant material in our Solar System (and potentially being more suitable for comparisons with dust from other planetary systems), OCs are more readily available for laboratory analysis due to their durability and higher recovery rates. 
    Our laboratory is working to build a spectral library, starting with Chilean samples from the Atacama Desert. According to the meteoritical database, Carbonaceous Ornans group (CO) are the most common types of CCs found in the Atacama Desert. Therefore, the samples used in this study consist of OCs and COs, with plans to expand the library to include a broader variety of meteorite types in order to be more representative of the diversity of material found in the Solar System and protoplanetary disks. 

    Meteorite classes are subdivided into clans, and then these clans are further divided into groups \citep{2006mess.book...19W}. A clan has similar chemical compositions, mineralogy, and O-isotopic content but different petrographic and/or bulk chemical characteristics, suggesting that material with similar petrogenesis likely come from the same reservoir, and thus the difference between two objects could be the product of a secondary process. In the case of OCs, the difference is not presented at the clan level but at the group level. A group is the basic classification unit and usually indicates meteorites from the same parent body \citep{2022M&PS...57.1774J}. They have similar petrologic (chondrule size, modal abundances, mineral composition), whole-rock chemical, and O-isotopic characteristics \citep{2006mess.book...19W}.  

    Besides chondrules, chondrites are made of other primary components and secondary minerals formed through geological processes: chondrule fragments; refractory inclusions such as calcium-aluminum rich inclusions (CAIs); olivine-rich aggregates; hydrated mineral assemblages -- for example, we can have Tochilinite/Cronstedtite Intergrowths (TCIs); mineral fragments (from collisional processes); and opaque minerals, and they are cemented in a fine-grained matrix, which may or may not include fine-grained rims surrounding components \citep{1992GeCoA..56.2873M, 2016M&PS...51..785P, 2019M&PS...54.1870V}.
    Primary mineralogy (pristine material) tends to be partially or almost completely altered through secondary geological processes.
    For instance, meteorites can experience (i) aqueous alteration, which leads to the formation of hydrated phases at the expense of the initial mineralogy (e.g., primary olivine and pyroxene are replaced by hydrated silicates, notably phyllosilicates), the oxidation of some metals into magnetite, and the formation of an aqueous alteration product; (ii) thermal metamorphism, which leads to dehydration, olivine recrystallization, and chemical equilibration; or (iii) collisional processes that can be recorded as brecciation, chondrule fractures, chondrule flattening, and/or potential impact heating \citep{1991GeCoA..55.3845S, 2007AREPS..35..577S, 2014mcp..book...65S, 2015GeCoA.148..159L, 2021AmMin.106.1388H}. Aqueous alteration and thermal metamorphism are often interlinked (e.g., the melting of ice caused by heating leads to aqueous alteration). These processes may result from preaccretionary and/or postaccretionary events, such as the melting of ices inside the parent body, radiogenic heat, solar and/or collisions/impacts, among others \citep{2007AREPS..35..577S, 2009GeCoA..73.4963K, 2014mcp..book...65S, 2021AmMin.106.1388H, 2024SSRv..220...69M}. 

    Furthermore, it is important to consider that chondrites are assigned a number according to the petrologic type defined by \citet{1967GeCoA..31..747V}, which was later expanded by several studies \citep{1990GeCoA..54.2485S, 2005M&PS...40...87G, 2007GeCoA..71.2361R, 2012M&PSA..75.5297A, 2013GeCoA.123..244A}. This number indicates a relation between pristine and altered material, where type 3.0 is ideally meant to represent the most pristine material, types 3.1 to 6 indicate an increasing degree of thermal metamorphism, and types 2 to 1 represent a rising degree of hydrous alteration. 
    In this way, different meteorites provide information from various stages of planet formation. Some meteorites (e.g., CCs) contain pristine material from the early Solar System, including CAIs, which are as old as the Sun, 4.56 billion years \citep{2005GeCoA..69.5805K, 2018GeCoA.228...62S}. 

    Undifferentiated meteorites, such as OCs, trace back to the early stages of disk evolution in the solar nebula, which is analogous to the Class I stage of young stellar objects, when disks are still embedded in their natal envelopes (see \cite{2011ARA&A..49...67W} for the classification of young stellar objects). However, even in the very early phases of protoplanetary disks, dust can undergo transformative processes, including thermal metamorphism, collisions, and fragmentation. In contrast, achondrites are derived from larger parent bodies that have already undergone differentiation \citep{gupta2010differentiation, 2012GeCoA..85..142G, 2015GeCoA.168..280B, 2015M&PS...50.1750D, 2020M&PS...55..857T}.   

    Several studies have already presented MIR absorption spectra of meteorites \citep{2008M&PS...43.1147M,
    2010Icar..207...45M,
    2012Icar..219...48M,
    2014Icar..231..338M, 
    2014Icar..229..263B,
    2018Icar..307..400D}.
    However, few works have studied the effect of particle size on observed spectra or measured their absolute mass absorption coefficients \citep{2006P&SS...54..599M}.  

    This study aims to present the first results of the Universidad Diego Portales (UDP) Cosmic Dust Laboratory, a long-term effort to contribute to the construction of dust spectral libraries for astronomical applications.  
    The first components of this library are the MIR (2 - 23 \textmu m) spectra and the mass absorption coefficient of the meteorite powder of known composition and particle size distributions. The sample includes 20 OCs and 3 CCs collected in the Atacama Desert and some pure minerals (olivine and diopside). In Sect. \ref{section_2}, we provide background information on the meteorite samples and describe the equipment, sample preparation methods, and laboratory measurements. In Sect. \ref{section_3}, we present the absorbance spectra and the mass absorption coefficients obtained. 
    In Sects. \ref{section_4} and \ref{section_5}, we discuss our results and some of their astronomical applications.
    A summary of our conclusions is presented in Sect. \ref{section_6}.

\begin{table*}
    \centering
    \caption{Official name of meteorites from the Meteoritical Bulletin Database.}
    \label{tbl:sample}
        \begin{tabular}{lllcl}
            \noalign{\smallskip}
            \hline
            \noalign{\smallskip}
            \textbf{Meteorite name}&\textbf{Abbreviation}&\textbf{Location}&\textbf{Year}&\textbf{Data Base} \\ 
            \noalign{\smallskip}
            \hline
            \noalign{\smallskip}
            Catalina 008    & C008   & Anf., Chile & 2011 & MB102 (2013) \\
            Los Vientos 123 & LoV123 & Anf., Chile & 2015 & MB105 (2016) \\
            El Médano 216   & EM216  & Anf., Chile & 2011 & MB103 (2014) \\
            \noalign{\smallskip}
            \hline
            \noalign{\smallskip}                                                        
            La Yesera 001           &        & Anf., Chile & 2003 & MB88 (2004) MetBase (2006) \\
            SanJuan02               & SJ002  & Anf., Chile & 2002 & MB87 (2003) MetBase (2006) \\        
            Estacion Imilac             &        & Anf., Chile & 2004 & MB100 (2012) \\                       
            Cobija                  &        & Anf., Chile & 1892 & NHM catal. 5th Ed. (2000) MetBase (2006) \\        
            Rencoret 001            &        & Anf., Chile & 1996 & MB101 (2012) \\                       
            Pampa de Mejillones 002 & PdM002 & Anf., Chile & 2003 & MB100 (2012) \\       
            \noalign{\smallskip}
            \hline
            \noalign{\smallskip}                                                        
            Pampa de Mejillones 004 & PdM004 & Anf., Chile    & 2003 & MB100 (2012) \\       
            Pampa de Mejillones 007 & PdM007 & Anf., Chile    & 2003 & MB100 (2012) \\       
            La Yesera 003           &        & Anf., Chile    & 2003 & MB100 (2012) \\       
            La Yesera 004           &        & Anf., Chile    & 2003 & MB100 (2012) \\       
            La Yesera 002           &        & Anf., Chile    & 2003 & MB88 (2004) MetBase (2006) \\
            Pampa de Mejillones 010 & PdM010 & Anf., Chile    & 2004 & MB100 (2012) \\       
            Pampa de Mejillones 011 & PdM011 & Anf., Chile    & 2004 & MB100 (2012) \\       
            Pampa (a)               &        & Anf., Chile    & 1986 & MB65 (1987) NHM catal. 5th Ed (2000) MetBase (2006) \\
            Pampa (b)               &        & Anf., Chile    & 1986 & MB65 (1987) NHM catal. 5th Ed (2000)    MetBase (2006) \\
            Pampa (c)               &        & Anf., Chile    & 1986 & MB67 (1989) NHM catal. 5th Ed (2000)    MetBase (2006) \\
            Pampa (d)               &        & Anf., Chile    & 1986 & MB85 (2001) NHM catal. 5th Ed (2000) MetBase (2006) \\                               
            Pampa (g)               &        & Anf., Chile    & 2000 & MB85 (2001) NHM catal. 5th Ed (2000) MetBase (2006) \\                               
            San Juan 001            & SJ001  & Anf., Chile    & 2001 & MB87 (2003) MetBase v.7.1 (2006) \\  
            Lutschaunig's Stone     &        & Atacama, Chile & 1861 & NHM catal. 5th Ed. (2000) MetBase (2006) \\ 
            \noalign{\smallskip}
            \hline
            \noalign{\smallskip}
        \end{tabular}
            \begin{tablenotes}
            \small
            \item {This table presents the official name of each meteorite, the official abbreviation name, the location it was found, the year it was found, and the meteoritical bulletin in which it was registered.}
            \end{tablenotes}
\end{table*}

\begin{table*}
    \centering
    \caption{Classification and properties of previously measured meteorites presented in this work.}
    \label{tbl:shock-weathering}
        \begin{tabular}{llllllll}
        \noalign{\smallskip}
        \hline
        \noalign{\smallskip}    \textbf{Meteorite}&\textbf{Class}&\textbf{Shock}&\textbf{Weathering}&\textbf{Terrestrial}&\textbf{Fayalite}&\textbf{Ferrosilite}&\textbf{Wollastonite}\\
        \textbf{name}& &\textbf{stage}&\textbf{grade}&\textbf{ages (ka)}&\textbf{mol[\%]}&\textbf{mol[\%]}&\textbf{mol[\%]}\\
        \noalign{\smallskip}
        \hline
        \noalign{\smallskip}
        C008   & CO3   & & moderate &  & 19.1 $\pm$ 13.5 & 2.9 $\pm$ 1   & 5.5 $\pm$ 1.1.6       \\      
        LoV123 & CO3.1 & & low &  & 13.6 $\pm$ 18.7 & 15.1+-19.6    & 0.9 $\pm$ 0.4 \\    
        EM216  & CO3   & & low &  & 4.4 $\pm$ 3.2   & 1.5 $\pm$ 0.7 & 1.2 $\pm$ 0.2 \\
        & & & & & 39.3 $\pm$ 9.7 \\
        \noalign{\smallskip}
        \hline
        \noalign{\smallskip}                                            
        La Yesera 001   & H6 & S2 & W3 & 10.79 $\pm$ 1.56 & 18.2  & 17.7 \\                              
        SJ002           & H6 & S1 & W3 & 19.44 $\pm$ 1.69 & 19.25 & 19 \\ 
        Estacion Imilac & H5 & S4 & W1 &                  & 18.0  & 16.0 \\                      
        Cobija          & H6 &    & W1 & 19.7 $\pm$ 4.20 \\     
        Rencoret 001    & H6 & S3 & W3 & 25.3 $\pm$ 6.4   & 19.8 $\pm$ 1.1 \\                      
        PdM002          & H5 & S2 & W3 & 3.86 $\pm$ 1.36 & 18.3  & 16.9 \\      
        \noalign{\smallskip}
        \hline
        \noalign{\smallskip}                                                    
        PdM004        & L6   & S3   & W4/5 & 23.98 $\pm$ 4.40 & 25.6 & 21.6 \\      
        PdM007        & L6   & S3   & W4   & >27.9            & 24.8 & 21.2 \\      
        La Yesera 002 & LL5  & S2   & W2   & 25.44 $\pm$ 4.45 & 28.2 & 25.1 \\      
        La Yesera 003 & L4   & S3   & W4   & 16.98 $\pm$ 2.47 & 24.2 & 20.5 \\      
        La Yesera 004 & L6   & S2   & W3   & 34.07 $\pm$ 1.92 & 25.9 & 21.1 \\      
        PdM010        & L5   & S3   & W3   & 18.06 $\pm$ 1.90 & 25.1 & 21.8 \\      
        PdM011        & L5   & S4/5 & W5   &  4.35 $\pm$ 1.34 & 25.3 & 21.5 \\      
        Pampa (a)     & L6   & S4   & W2   & 25.08 $\pm$ 1.46 \\                                                                
        Pampa (b)     & L4/5 & S4   & W3   & 21.29 $\pm$ 2.45 \\                                                                
        Pampa (c)     & L4   & S5   & W4   & 13.89 $\pm$ 2.08 \\                                                                
        Pampa (d)     & L5   & S2   & W2/3 & 14.18 $\pm$ 1.91 & 25   & 22.9 \\                              
        Pampa (g)     & L5   & S2   & W3   & 14.34 $\pm$ 1.62 & 24.5 & 22.4 \\                              
        SJ001         & L5   & S1/2 & W2   & >28.1            & 24.5 & 21.6 \\      
        Lutschaunig's stone & L6 &  & W1   &  9.2 $\pm$ 1.6 \\  
        \noalign{\smallskip}
        \hline
        \end{tabular}
            \begin{tablenotes}
            \small
            \item {The classification, shock stage, and grade weathering are presented for each meteorite along with terrestrial ages (Ka) and olivine content for some of them. The data are from The Meteoritical Bulletin. Terrestrial ages are from \citep{valenzuela2011procesos}.}
            \end{tablenotes}
    \end{table*}

\begin{table*}
    \caption{Mineral abundances of some samples of ordinary chondrites characterized in this study as determined by X-ray diffraction.}
        \label{tbl:XRD}
        \begin{tabular}{lcccccccccccc}
        \hline
        \noalign{\smallskip}
        \textbf{Official}&\textbf{Class}&\textbf{Ol}&\textbf{Px}&\textbf{Fe-Ni}&\textbf{Tro}&\textbf{Ab}&\textbf{An}&\textbf{Mag}&\textbf{Hem}&\textbf{Gth}&\textbf{Akg}&\textbf{Total}\\ \textbf{Name}& &\textbf{[\%]}&\textbf{[\%]}&\textbf{[\%]}&\textbf{[\%]}&\textbf{[\%]}&\textbf{[\%]}&\textbf{[\%]}&\textbf{[\%]}&\textbf{[\%]}&\textbf{[\%]}&\textbf{[\%]}\\
        \noalign{\smallskip}
        \hline
        \noalign{\smallskip}
        PdM002           & H5 S2 W3  & 40.96& 31.05& 0.00& 2.80& 7.48& 4.71 & 4.92& 0.22& 5.51 & 2.36& 100.00 \\ 
        Est. Imilac      & H5 S4 W1  & 45.50& 31.92& 0.31& 2.43& 4.76& 10.49& 3.92& 0.00& 0.36 & 0.31& 100.00\\
        SJ002            & H6 S1 W3  & 32.37& 36.96& 0.00& 1.67& 6.97& 6.06 & 3.22& 0.20& 12.22& 0.34& 100.01\\ 
        Rencoret 001     & H6 S3 W3  & 40.60& 35.34& 0.00& 1.94& 1.72& 11.17& 4.73& 0.26& 2.83 & 1.40& 99.99\\ 
        La Yesera 003    & L4 S3 W4  & 44.56& 28.79& 0.00& 0.71& 5.28& 6.02 & 3.00& 0.22& 9.92 & 1.32& 99.82\\ 
        Pampa (b)        & L4/5 S4 W3& 48.81& 29.21& 0.00& 1.47& 3.02& 8.47 & 3.82& 0.00& 3.35 & 1.55& 99.70\\ 
        Pampa (d)        & L5 S2 W2/3& 41.48& 33.74& 0.00& 0.95& 3.52& 8.05 & 4.09& 0.20& 6.80 & 1.18& 100.01\\ 
        Pampa (g)        & L5 S2 W3  & 32.70& 37.09& 0.00& 0.00& 5.10& 7.23 & 3.44& 0.25& 14.15& 0.03& 99.99\\ 
        La Yesera 004    & L6 S2 W3  & 44.30& 35.11& 0.01& 0.00& 1.52& 7.49 & 3.38& 0.20& 6.07 & 1.02& 99.10\\ 
        PdM007           & L6 S3 W4  & 36.75& 23.74& 0.00& 0.00& 4.70& 14.48& 8.44& 0.24& 11.65& 0.00& 100.00\\ 
        PdM004           & L6 S3 W4/5& 43.19& 27.19& 0.00& 0.00& 3.01& 11.62& 4.25& 0.33& 10.40& 0.00& 99.99\\ 
        Lutschaunig stone& L6 W1     & 55.01& 28.85& 0.18& 2.94& 5.77& 1.66 & 4.58& 0.00& 0.78 & 0.23& 100.00\\ 
        La Yesera 002    & LL5 S2 W2 & 50.95& 27.14& 0.00& 1.28& 3.09& 13.12& 1.58& 0.20& 2.69 & 0.00& 100.05\\ 
        \noalign{\smallskip}
        \hline
        \end{tabular}
        \begin{tablenotes}
            \small
            \item {These data are from \citep{valenzuela2011procesos}. Abundances are presented in wt[\%]. Pyroxene corresponds to diopside and enstatite. Primary minerals are olivine, pyroxene, Fe-Ni, and troillite. Secondary minerals minerals are formed from secondary processes, including both thermal metamorphism and aqueous alteration (albite, anorthite, and likely magnetite). Minerals formed from terrestrial weathering are separate from these (goethtite, hematite; IMA symbol).}
            \end{tablenotes}
    \end{table*}

\section{Laboratory measurements}\label{section_2}
\vspace{0.3cm}
\subsection{Samples}
\begin{figure}[]
        \centering
                \includegraphics[width=0.4\textwidth]{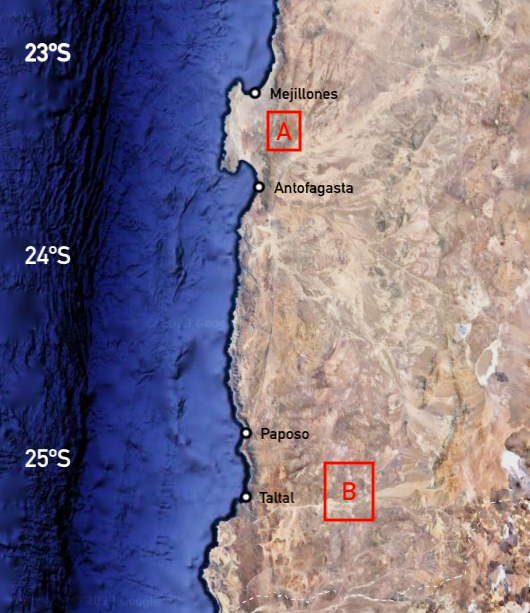}
          \caption{Location of the collected samples from the Atacama Desert in the Antofagasta Region, Chile. A) Pampa de Mejillones, southern of Mejillones city. B) Areas of San Juan, Catalina, Los Vientos, and El Médano, inland the city of Taltal. The CO meteorites come from this area. The one exception is Lutschaunig’s stone, which cannot be precisely located on this map, as no records exist regarding its specific discovery location. }\label{fig:mapa}
\end{figure}

\vspace{0.3cm}
    The samples studied herein include 23 stony meteorites collected in the Atacama Desert in the Antofagasta region, northern Chile. 
    Following the standard classification \citep{2004mete.book.....H, 2006mess.book...19W}, and according to the meteoritical database, three of the meteorites in the sample correspond to Ornans carbonaceous chondrites (CO group) from the areas of "Catalina", "Los Vientos" and "El Medano", inland Taltal. 
    The remaining twenty correspond to OCs of three different groups, classified according to the abundance of Fe and the ratio of metallic Fe (Fe\textsuperscript{0}): H chondrites with Highest total iron, high Fe-Ni metal and lower iron content in silicate minerals (Fe\textsuperscript{0}/FeO = 0.58), L chondrites with Low total iron, low Fe-Ni metal and higher iron content in silicate minerals (Fe\textsuperscript{0}/FeO = 0.29) and LL chondrites with Low total iron and Low Fe-Ni metal and highest iron content in silicates minerals \citep[Fe\textsuperscript{0}/FeO = 0.11;][]{2006mess.book...19W}. They were collected from the Pampa de Mejillones area, a plain deflation zone of $\sim$ 250 km\textsuperscript{2} in the Mejillones peninsula, and from different deflation surfaces in the Central Depression $\sim$ 100 km east of Taltal (Fig.\ref{fig:mapa}). The one exception is Lutschaunig's stone, which has no specified region in the Atacama Desert (see Table \ref{tbl:sample}).

    All meteorites that we studied are registered in The Meteoritical Bulletin, and we present data on the location and year in which they were found in Table \ref{tbl:sample}. The classification, shock stage, weathering information, magnetic susceptibility, and olivine content are available in Table \ref{tbl:shock-weathering}. Finally, chemical analysis and mineralogical characterization from \citep{valenzuela2011procesos} are available in Table \ref{tbl:XRD}.

\subsection{Sample preparation}
\vspace{0.3cm}
    Originally, our meteorite samples were in powder format and the mass available for each meteorite ranged from 20 mg to 50 mg. The initial grain size distribution varied from meteorite to meteorite. We first measured each meteorite with the original (unsieved) size to obtain a general spectrum of the sample. We used a dry agate mortar to create a mixture of sample-powder and ultrapure KBr (potassium bromide, commercial IR grade Uvasol). We prepared a solid pellet by pressing the powder mixture with a Specac hydraulic press at 1.7 tons for 5 minutes. Then we measured the prepared solid pellet with a mass concentration ratio of 0.5 \% to 1 \% (sample/KBr).  

    In order to obtain smaller grains, we grounded the mixture in the agate mortar for 60 - 75 minutes by hand. The mixture was dried for 24 hours at 110°C in a natural convection oven to guarantee water evaporation and minimize terrestrial or ambient water contamination. The dry mixture powder is compressed at 1.7 tons for 3 minutes to obtain a 7 mm pellet with optimal optical quality. The compression process does not affect the grain size distribution of the sample (see Sect. \ref{absgrainsize}). Each pellet was weighed in a microbalance (0.1 mg sensitivity) to register the sample mass.

\subsubsection{Sample sieving}
\vspace{0.3cm}    
    The small amount of material available for the meteorite samples prevented us from using them to explore the effect of larger grain sizes on the absorbance. However, we were able to perform the experiment using pure minerals: olivine and diopside. We used the sieves to produce samples with grain sizes in the following ranges: G) 300 – 250 \textmu m, F) 250 – 125 \textmu m, E) 125 – 75 \textmu m, D) 75 – 63 \textmu m, A) 45 – 38 \textmu m, lA) $<$ 38 \textmu m.  We note that increasing the grain size also requires increasing the concentration of the samples to a few percent by mass in the KBr pellets. 
    The adopted nomenclature for grain sizes is listed in Table \ref{tbl:meshsizes}. For instance, sample A-B corresponds to grains that went through sieve-B with a mesh size of 45 \textmu m but did not go through sieve-A with a mesh size of 38 \textmu m.

\begin{table}
    \centering
    \caption{Mesh size and nomenclature used in this work.}
        \label{tbl:meshsizes}
        \begin{tabular}{p{0.5\linewidth}cc}
        \hline
        \noalign{\smallskip}
        \textbf{Nomenclature}&\textbf{Particle size range [\textmu m]}\\ 
        \noalign{\smallskip}
        \hline
        \noalign{\smallskip}
        A        &         <38 \\
        A-B      &      38 - 45 \\
        B-C      &      45 - 53 \\
        C-D      &      53 - 63 \\
        D-E      &      63 - 75 \\
        E-F      &      75 - 125 \\
        F-G      &      125 - 250 \\
        G-H      &      250 - 300 \\
        \noalign{\smallskip}
        \hline
        \end{tabular}
    \end{table}

\subsection{Spectroscopic measurements}
\vspace{0.3cm}
    Transmission infrared spectra of samples were measured with the experimental setup of the UDP Cosmic Dust Laboratory in Santiago, Chile. The laboratory is equipped with a Bruker Vertex 80v Fourier transform infrared (FTIR) spectrometer.
    The spectra were measured using 
    a HeNe laser, 
    a KBr beamsplitter,
    a Globar IR light source, 
    and a DLaTGS detector, with a resolution of 4 cm\textsuperscript{-1}. 
%
    The measurements were performed at ambient temperature, using a KBr window, and at primary vacuum (P = 1 - 3 mbar). 
    The spectral range is from 2 \textmu m to 23\textmu m, with a standard error of 0.01 \textmu m.
    Each final spectrum is obtained by averaging sixteen measurements. Furthermore, we measured a background spectrum of pure KBr pellet, with the sample compartment empty and under vacuum $\sim$ 1hPa, which is subtracted from each measurement to obtain the absorption spectrum of the samples with the least possible influence of KBr.

\subsection{Absorbance and opacities}
\vspace{0.3cm}
    The transmittance spectrum of each sample is obtained directly by dividing the spectrum of the sample with KBr, by the spectrum of a blank pellet with pure KBr, which is mostly transparent in the infrared range. 
    The Mass Absorption Coefficient (MAC) of the sample, $\kappa_{sample}$ (cm\textsuperscript{2}g\textsuperscript{-1}), is calculated from the transmittance spectra following the formula of \citet{1998asls.book.....B}:
\begin{gather}
    \kappa_{\text{sample}} = - \frac{S}{M} \times \log{(T)},
\end{gather}

\noindent where S is the transverse section of the pellet (area in cm\textsuperscript{2}), M is the mass of the sample in the pellet (g) and T is the transmittance spectrum. The absorbance is related to the transmittance by the equation
\begin{gather}
    A = \log{\left(\frac{1}{T}\right)}.
\end{gather}

\subsection{Microphotography}\label{Micro-photography}
\vspace{0.3cm}
    In order to characterize the particle size distribution of the samples, we applied image analysis to microphotography.
    The microphotography images were taken after the spectroscopic measurements. 
    To prepare the samples for photography, we put each pellet in distilled water in a vial, and the dissolved sample is shaken in an ultrasonic bath for 15 seconds to properly disperse the grains and avoid conglomerates.

    Each microphotography was taken using a Microscope Leica LM400, with transmitted light with two modes: bright field and dark field, with the largest number of grains possible for optimal statistics and with different amplification (10x, 20x, 50x, and 100x).
    The images were acquired with a Neubauer camera. Reference images were taken for each amplification using reference squares of 100 \textmu m, 50 \textmu m, and 5 \textmu m to build a real scale for pixel and obtain the grain size distribution for each image.

    The images were processed with the open-source software ImageJ \footnote{https://imagej.net} to obtain the particle size distribution of each sample.
    A representative photograph of one chondrite is shown in Sec. \ref{absgrainsize}. Similar photographs were taken for all the samples. 

\section{Results}\label{section_3}
\vspace{0.3cm}
    In this section we present the absorbance and opacity spectra of the samples. Regarding absorbance, we focus on the shape and relative intensity of the peaks around 10 \textmu m and 20 \textmu m to observe the silicate features. As detailed in the following analysis, the most common minerals in meteorites are olivine and pyroxene. These two minerals are a solid solution, with a range of compositions with two different extreme members, ranging from Mg-rich to Fe-rich. Olivines (Mg$_{2x}$Fe$_{2(1-x)}$SiO$_{4}$) range from fayalite (x = 0, iron-rich endmember) to forsterite (x = 1, Mg-rich endmember), and pyroxenes (Mg$_{x}$Fe$_{1-x}$Si$_{2}$O$_{6}$) range from ferrosilite (x = 0, iron-rich endmember) to enstatite (x = 1, Mg-rich endmember). 
    Regarding opacity, we are interested in absolute values that can be useful for radiative transfer modeling.

\subsection{Absorbance}
\begin{figure}[ht]
        \centering
                \includegraphics[width=6.65cm]{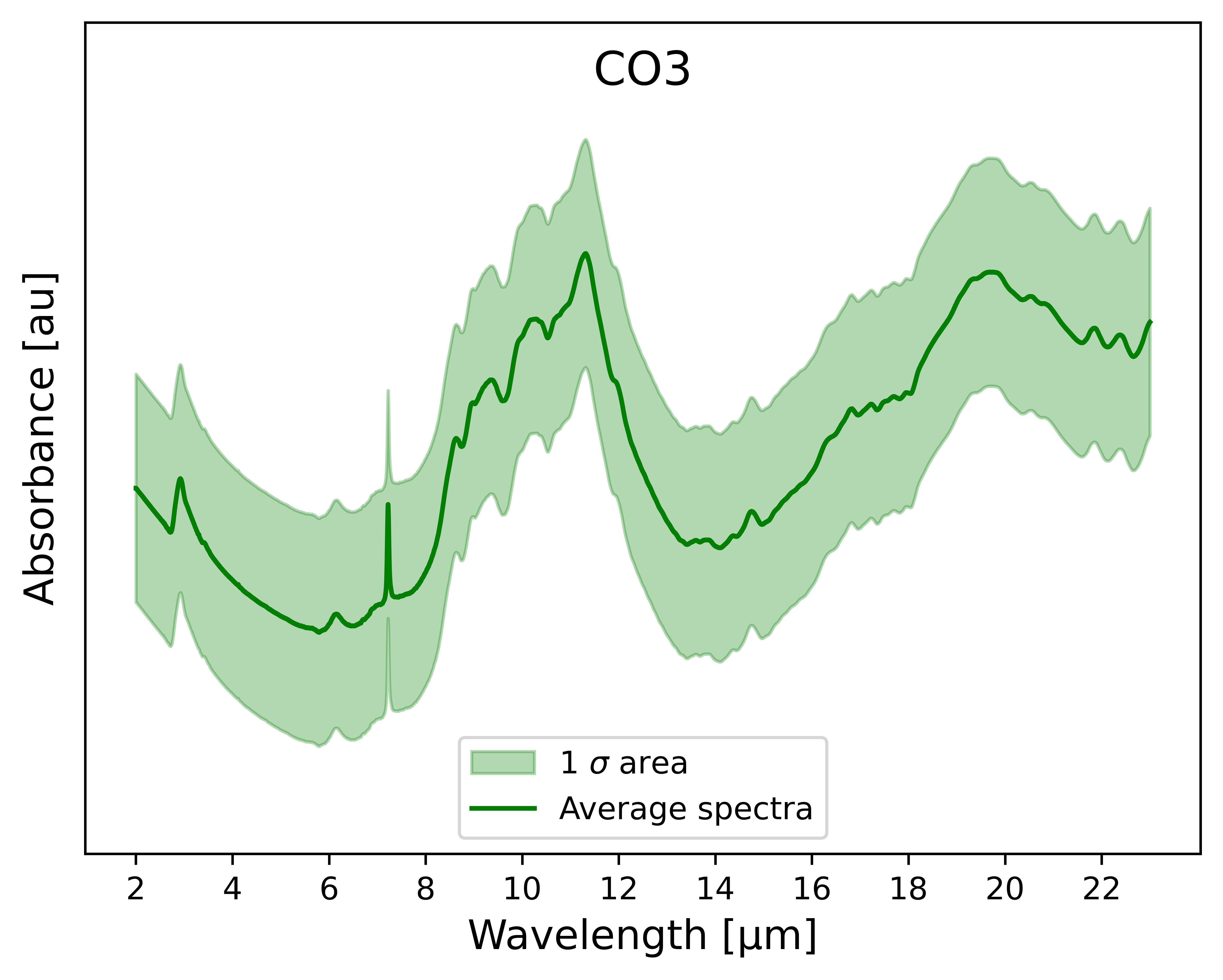}
        \includegraphics[width=6.65cm]{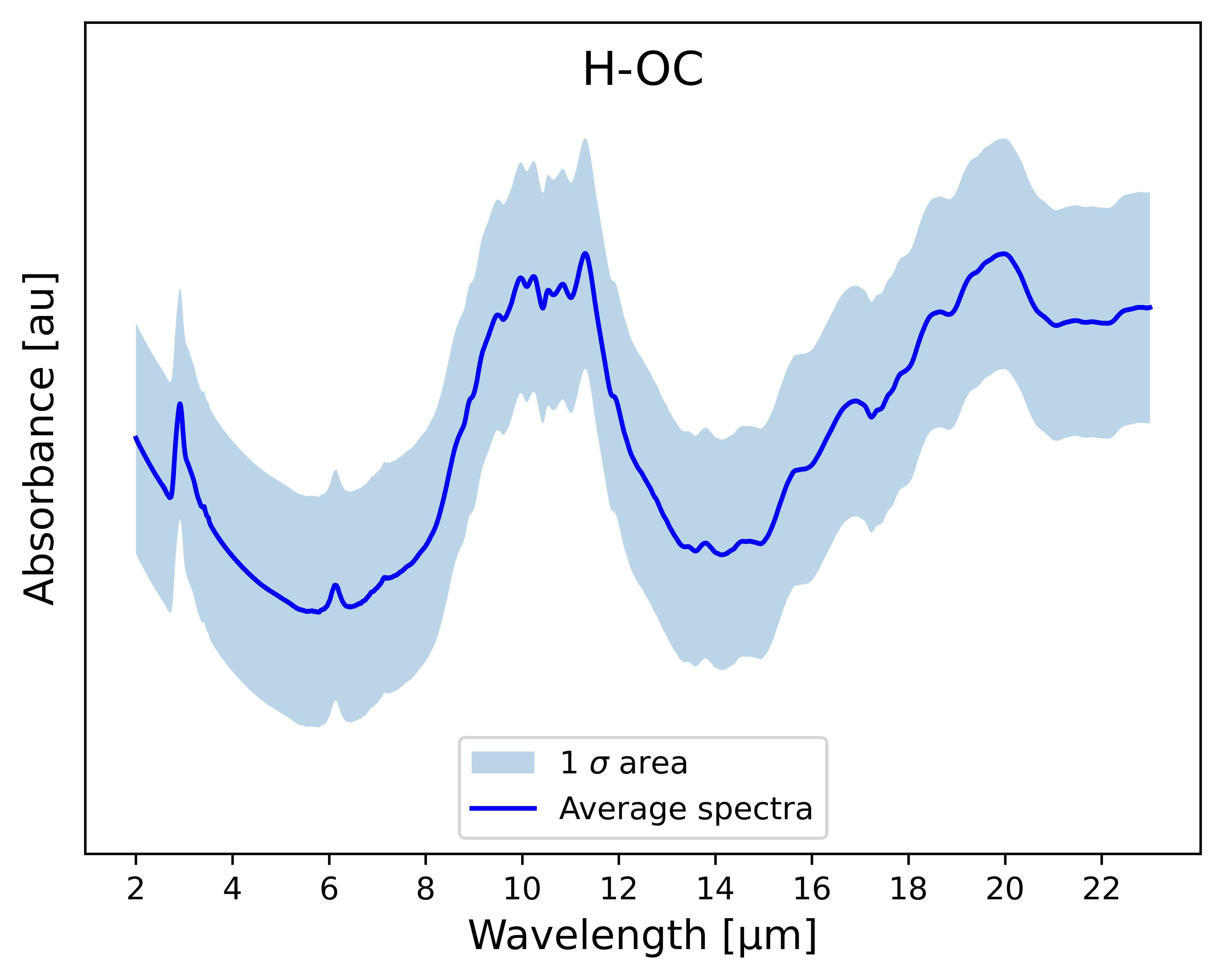}
        \includegraphics[width=6.65cm]{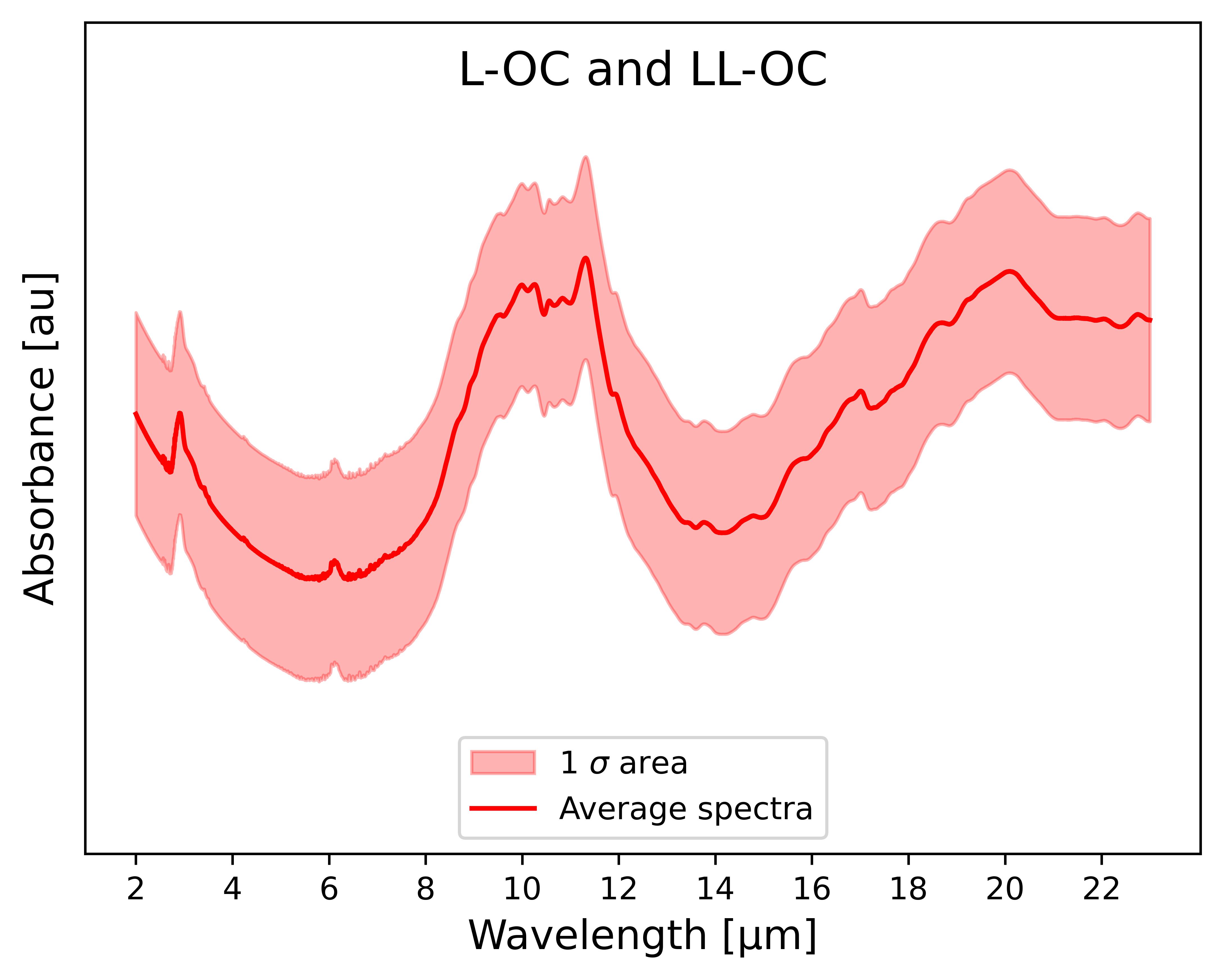}
        \includegraphics[width=6.65cm]{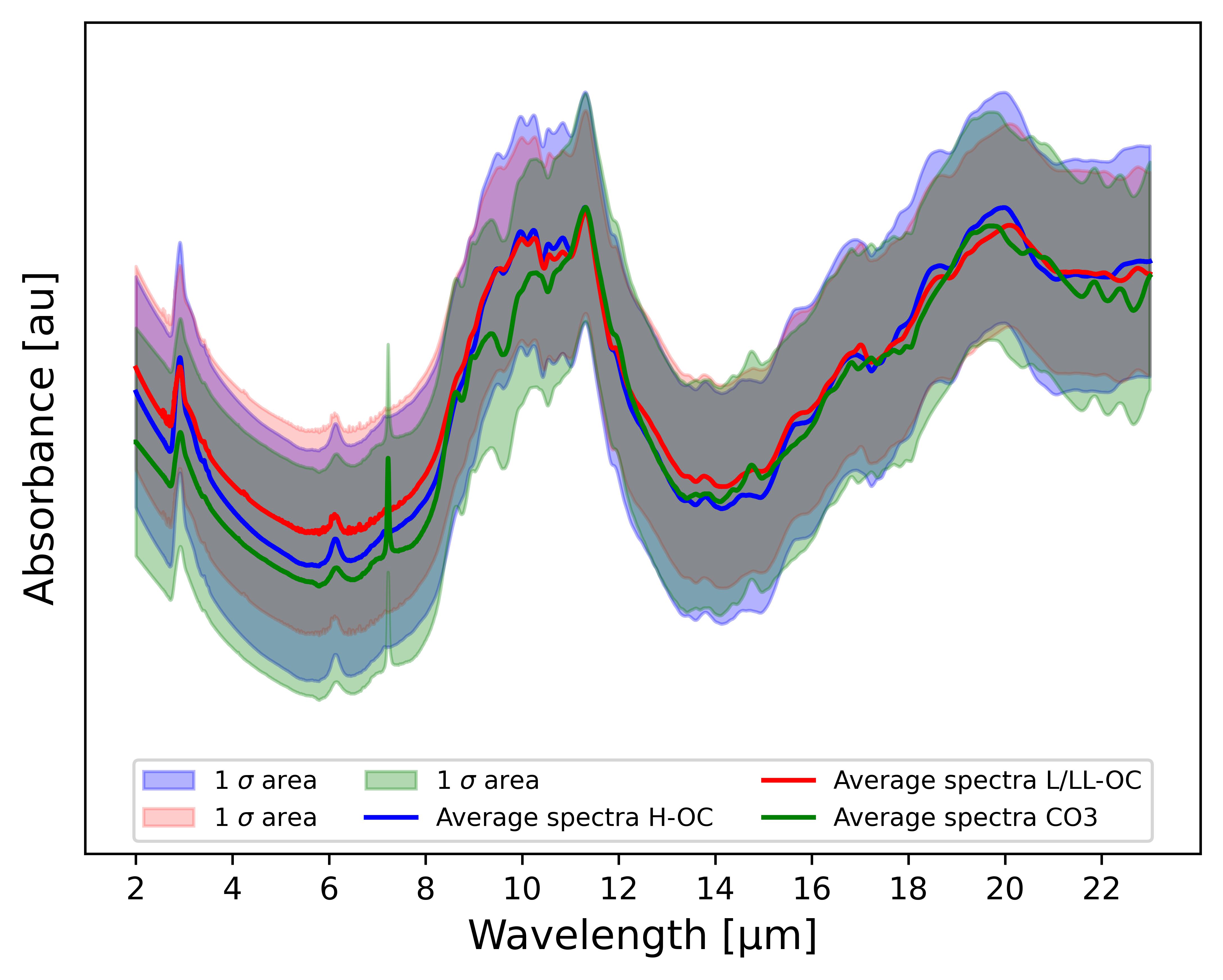}
            \caption{Average spectra for each group of meteorites in this study. The colored area is the confident interval for 1 $\sigma$.}
        \label{fig:overlap}
\end{figure}

\vspace{0.3cm} 
    In Fig. \ref{fig:overlap} we present a representative averaged spectrum of all spectra measured for each group of meteorites. 
    In the three groups, the $\sim$10 - 25 \textmu m range is dominated by vibrations of SiO$_4$, particularly features around 10 \textmu m are dominated by stretching modes of Si-O. Absorption bands around 3 \textmu m and 6 \textmu m are due to OH-bearing minerals, -OH stretching, X-OH bending, and a combination of both, and 12-16 \textmu m due to -OH vibration \citep{2002ApJ...566L.113G, 2005MNRAS.358.1383B, 2014Icar..229..263B}. These features can be used to quantify the level of hydration and degree of weathering of samples related to the classification by petrological description. However, this is beyond the scope of this work; therefore, we did not analyze the 3 \textmu m region at all, regardless of whether any features could be attributed to weathering.

    From this average, we observe a recognizable difference in the range of $\sim$10 \textmu m between the two main groups: CO and H-L-LL chondrites. Among OCs, distinguishing differences at the clan level is challenging due to their similar petrological and chemical characteristics. As shown in Table \ref{tbl:XRD}, OCs share a comparable mineralogy, consisting of approximately 70-80\% olivine and pyroxene. Among our CO3 samples, there are no significant variations; however, \cite{2021M&PS...56.1758P} studied the MIR spectra of CO3 chondrites and found significant variations. This could be correlated with their petrological grade. Based on the values of Cr$_2$O$_3$ content in Fe-rich olivine listed in The Meteoritical Bulletin, which can be used to classify petrologic grade type-3 chondrites \citep{2005M&PS...40...87G, 2019ChEG...79l5528R}, our three CO3s are all possibly of petrologic grade $\leq$ 3.2. The values also indicate that Los Vientos 123 is the least heated among the three.

\subsubsection{Absorbance of CO3 chondrites}
\begin{figure*}[ht]
        \centering
                \includegraphics[height=10cm]{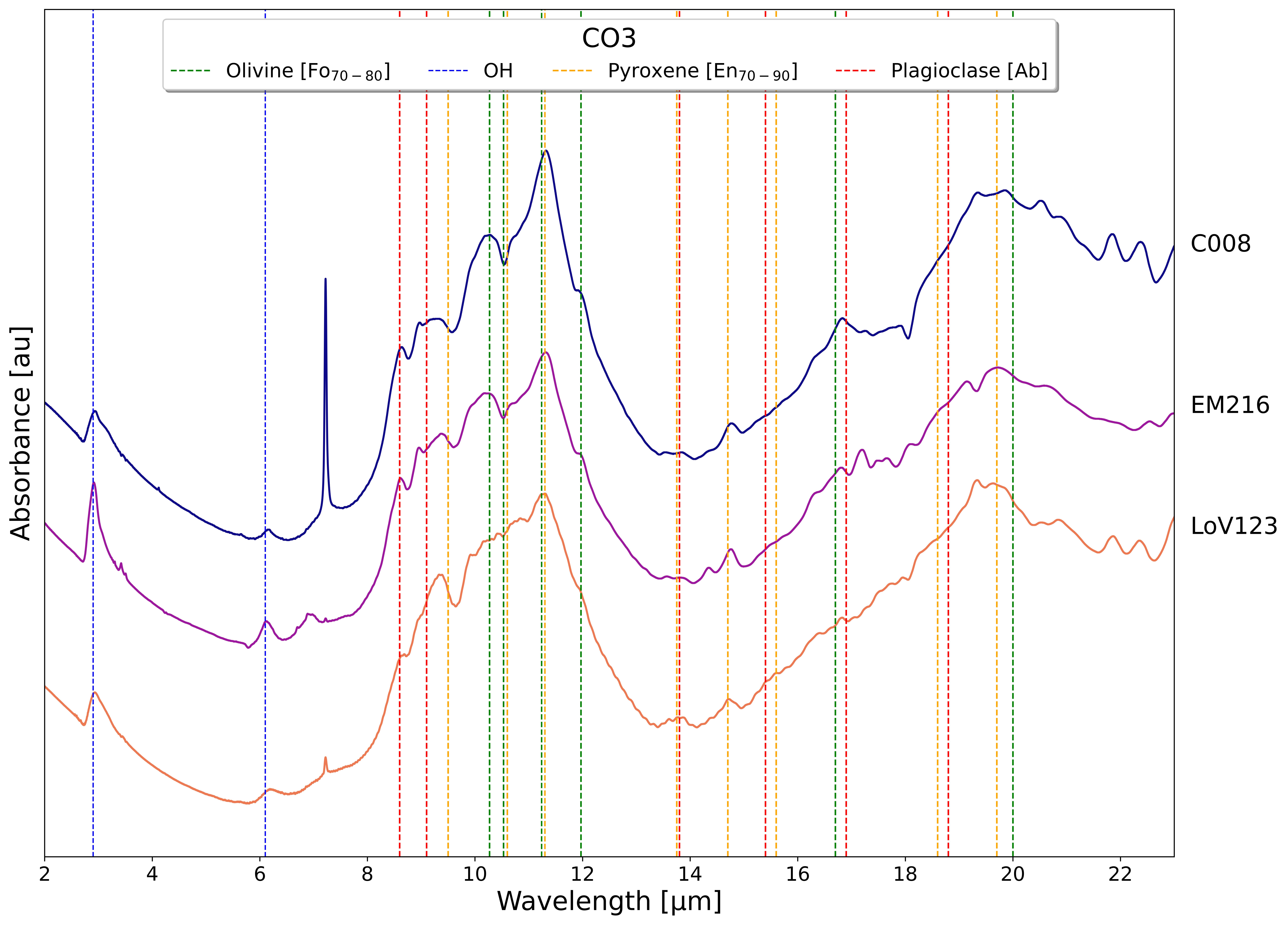}
            \caption{Mid-infrared spectra of CO3s. The spectra correspond to milled samples, with grain size between 1 - 10 \textmu m. The data presented are normalized to the maximum in the range of 8 - 12 \textmu m and vertically shifted for a better appreciation of shape and peaks. The vertical lines correspond to characteristic absorption bands of different minerals: light blue, for the water bands (OH); green for olivine bands, orange for the pyroxene bands, and red for plagioclase \citep{2002A&A...391..267C, 2003A&A...399.1101K, 2014Icar..229..263B, 2017P&SS..149...94C}.}
        \label{fig:abs-co}
\end{figure*}

\vspace{0.3cm} 
    The CO3s (Fig. \ref{fig:abs-co}) show very similar spectra, with the greatest difference in the relative intensity of the peaks around 10 \textmu m. 
    The peaks at 10.2, 11.3, 11.9, 16.3, and 19.7 \textmu m correspond to olivine. The sharpest peaks at 10.2 and 19.7 \textmu m should correspond to forsterite-rich olivine, Mg-rich \citep[Fo>60;][]{2003A&A...399.1101K, 2007PCM....34..319H, 2010MNRAS.406..460P}.
    Around 13-15 \textmu m, we see small features related to peaks at 9.5, 14.7, and 19.7 \textmu m indicating the presence of intermediate pyroxene \citep[En$_{50}$;][]{2000A&A...363.1115K, 2002A&A...391..267C, 2007MNRAS.376.1367B}. However, the sharpest feature around 10 \textmu m points to a predominance of olivine over other minerals in the samples.

    The LoV123 sample shows more peaks and features around 10 \textmu m that could imply the presence of different types of silicate in addition to olivine and pyroxene. These features may correspond to minerals such as plagioclase, which can be attributed to peaks at  [8.7, 10, 10.9 and 15.4] \citep{2017P&SS..149...94C}, or to diopside pyroxene, with peaks at [9.2, 10.41, 10.86, 15.8 and 18.5] \citep{2008PCM....35..399E, 2014isms.book.....C}. Similarly, the other samples, El Médano 216 and Catalina 008 show peaks at 8.7 \textmu m and 20.8 \textmu m that could be attributed to the presence of feldspar albite-anorthite \citep{2017P&SS..149...94C}.

    Likewise, in the three meteorite samples, a strong and well-defined (sharp and very narrow) peak is observed at 7.2 \textmu m, particularly in Catalina 008, where the relative intensity is greater. This band could appear as a product of carbonaceous material \citep{2001A&A...376..254K, 2002ApJ...566L.113G, 2004A&A...416..165R}, carbonates, or even sulfates.

    The peaks of each MIR spectrum are shown in Table \ref{tbl:abs-carbon}.

\begin{table}
    \centering
    \caption{Peak positions (in \textmu m) for the absorption bands in our three CO3 chondrite spectra.} \label{tbl:abs-carbon}
        \begin{tabular}{ccc}
        \noalign{\smallskip}
        \hline
        \noalign{\smallskip}
        \textbf{C008}&\textbf{LoV123}&\textbf{EM216}\\
        \noalign{\smallskip}
        \hline
        \noalign{\smallskip}
        2.92  & 2.92  & 2.95 \\
              &       & 3.07 \\
        3.42  & 3.42  & 3.42 \\
        3.50  & 3.50  & 3.51 \\
              & 4.23  & 4.24 \\
              & 4.28  & 4.29 \\
        6.12  & 6.11  & 6.19 \\
        6.72  & 6.97  & \\
        7.22  & 7.22  & 7.22 \\
        7.48  & 7.61  & 7.57 \\
        8.64  & 8.63  & 8.69 \\
              & 8.99  & \\
        9.35  & 9.37  & 9.38 \\
              &       & 9.95 \\
        10.24 & 10.23 & \\
              &       & 10.45 \\
              &       & 10.89 \\
        11.32 & 11.30 & 11.30 \\
              &       & 12.93 \\                
              & 13.21 & \\
              &       & 13.40 \\
        13.65 & 13.62 & \\
        13.81 & 13.80 & 13.84 \\
              & 14.39 & 14.37 \\
        14.76 & 14.76 & 14.71 \\
              &       & 15.49 \\
              & 16.40 & 16.32 \\
        16.82 & 16.82 & 16.72 \\
        17.27 & 17.21 & 17.29 \\
        17.77 & 17.68 & 17.55 \\
              &       & 18.04 \\
              &       & 18.27 \\
              &       & 18.62 \\
              & 19.16 & 19.15 \\
        19.77 & 19.80 & 19.71 \\
        20.60 & 20.61 & 20.82 \\
              & 21.62 & 21.89 \\
              &       & 22.38 \\
        \noalign{\smallskip}
        \hline
        \end{tabular}
    \end{table}

\subsubsection{Absorbance of OC type H}
\begin{figure*}[ht]
        \centering
                \includegraphics[height=10cm]{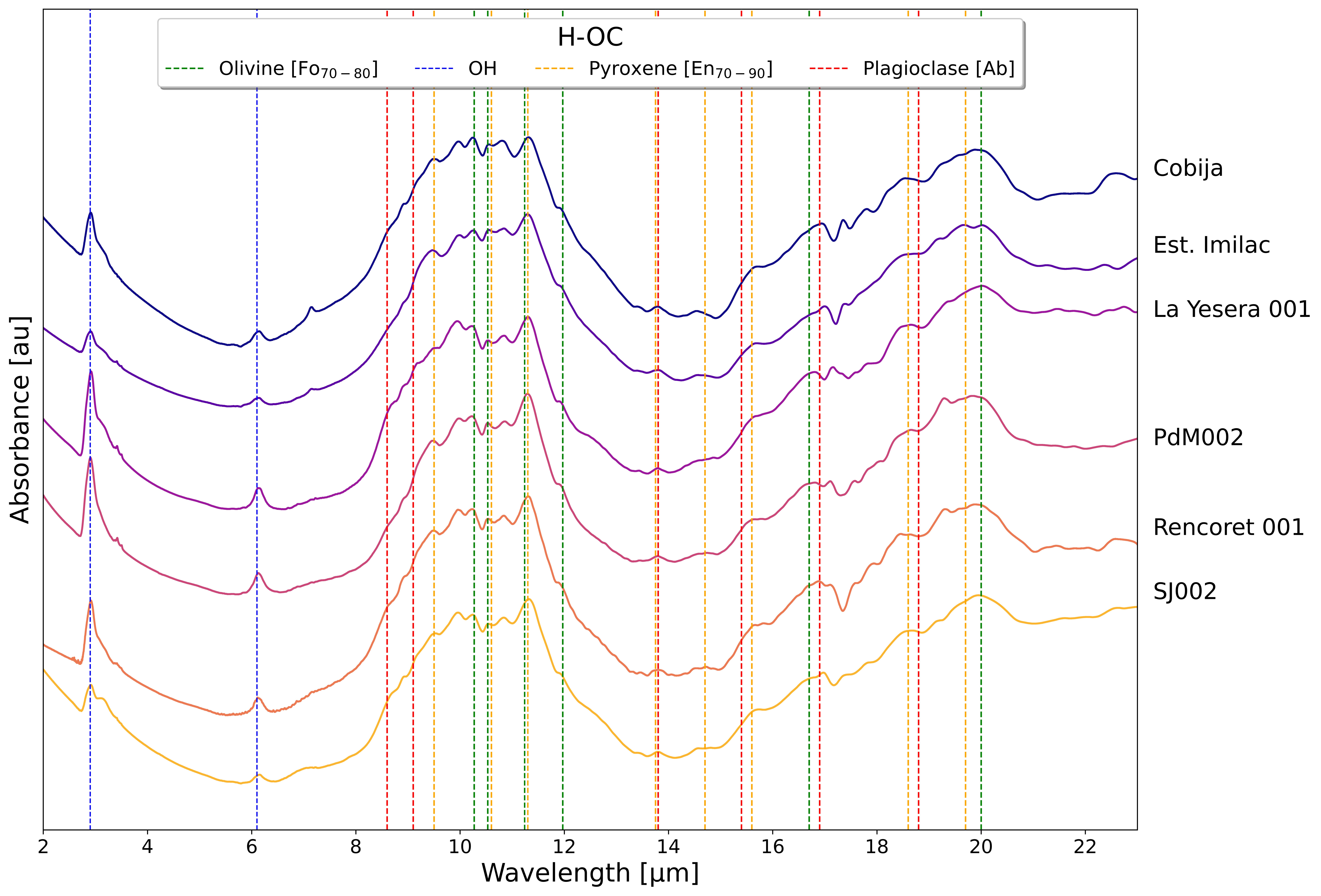}
            \caption{Mid-infrared spectra of H-OC. The spectra correspond to milled samples, with grain size between 1 - 10 \textmu m. The data presented are normalized to the maximum in the range of 8 - 12 \textmu m and vertically shifted for a better appreciation of shape and peaks. The vertical lines correspond to characteristic absorption bands of different minerals: light blue, for the water bands (OH); green for olivine bands, orange for the pyroxene bands, and red for plagioclase \citep{2002A&A...391..267C, 2003A&A...399.1101K, 2014Icar..229..263B, 2017P&SS..149...94C}.}
        \label{fig:abs-H}
\end{figure*}

\vspace{0.3cm}
    The spectra of the six ordinary chondrites belonging to group H (H-OC) are also very similar to each other (see Fig. \ref{fig:abs-H}), where the observed difference corresponds mainly to variations in the relative intensity of certain absorption bands.
    In comparison with the CO3 spectra, the H-OC spectra present a complex shape with several peaks in the 10 \textmu m area. These complex peaks indicate a mixture of silicates dominated mainly by sharp and very well-defined peaks at 10.2 and 11.3 \textmu m, which correspond to the presence of olivine. The absorption bands at 11.9 and 19.9 \textmu m also correspond to olivine. The slight differences in relative intensity between 10.2 and 11.3 \textmu m, and small shifts in the olivine peaks, are the product of variations in the composition of this mineral within the samples, which moves within the solid solution of forsterite-fayalite showing a predominance of forsteritic olivine \citep[Fo>50;][]{2003A&A...399.1101K, 2007PCM....34..319H, 2010MNRAS.406..460P}.

    The small peaks between 13 and 15 \textmu m are accentuated with respect to the CO3s chondrites. This, together with the peaks at 9.5, 10.3, 10.7, 15.7, 18.6, and 19.8 \textmu m, is associated with pyroxene enstatite. The small variations in the position of these bands are a product of the variation in the crystallography of the mineral: orthoenstatite (orthorhombic) and clinoenstatite (monoclinic), and variations within the composition of the solid solution enstatite - ferrosilite \citep{2000A&A...363.1115K, 2002A&A...391..267C, 2007MNRAS.376.1367B}.

    Although the peak at 8.7 \textmu m does not stand out in the spectra, along with the peaks at 9.9, 10.9, 16.0, 17.2, and 18.5 \textmu m indicates the presence of plagioclase feldspar. The slight variations in the position of these peaks correspond to differences in the composition within the solid solution of plagioclase albite – anorthite \citep{2017P&SS..149...94C}.

    Some absorption bands may be associated with multiple minerals, such as the one at 9.8 \textmu m, which indicates the presence both albite and enstatite and is consistent with other peaks in the spectrum. Similarly, the peak at 15.6 \textmu m suggests the presence of enstatite. Additionally, the bands at 15.6 and 18.5 \textmu m correspond to pyroxene enstatite and feldspar albite. The details of the peaks present in the MIR spectra of type H chondrites are provided in Table \ref{tbl:abs-H}.

\begin{table}
    \centering
    \caption{Peak positions (in \textmu m) for the absorption bands in our six H-OC spectra.}
    \label{tbl:abs-H}
    \small
        \begin{tabular}{cccccc}
        \noalign{\smallskip}
        \hline
        \noalign{\smallskip}
        \textbf{PdM}&\textbf{Cobija}&\textbf{Estacion}&\textbf{La Yesera}&\textbf{Rencoret}&\textbf{San Juan}\\
        \textbf{002}& &\textbf{Imilac} & \textbf{001}&  \textbf{001}& \textbf{002}\\
        \noalign{\smallskip}
        \hline
        \noalign{\smallskip}
        2.91 &2.91  &2.91  &2.92  &2.91  &2.92 \\
             &      &      &      &      &3.09 \\
             &      &      &3.38  &3.37  & \\
        3.42 &      &3.41  &      &3.41  &3.41 \\
        3.50 &      &      &3.50  &      & \\
             &5.66  &      &      &      &5.71 \\
        6.13 &6.15  &6.15  &6.14  &6.12  &6.16 \\
             &7.16  &7.16  &      &7.48  &7.07 \\
        9.49 &9.50  &9.48  &9.51  &9.50  &9.54 \\
        9.98 &9.98  &10.00 &9.96  &9.96  &9.97 \\
        10.23&10.25 &10.26 &10.23 &10.24 &10.25 \\
        10.53&10.57 &10.56 &10.54 &10.55 &10.55 \\
        10.87&10.81 &1084  &10.85 &10.85 &10.85 \\
        11.30&11.32 &11.30 &11.31 &11.32 &11.32 \\
             &      &      &11.90 &      & \\
             &13.40 &13.40 &13.44 &13.47 &13.40 \\
        13.79&13.81 &13.82 &13.80 &13.82 &13.80 \\
             &14.56 &14.57 &      &14.57 &14.57 \\
        14.75&      &      &14.88 &14.73 & \\
        15.64&      &      &      &15.67 & \\
             &15.75 &15.72 &      &      &15.74 \\
        15.81&      &      &      &14.84 & \\
        16.80&16.94 &17.01 &16.80 &16.90 &16.98 \\
        17.11&17.35 &17.41 &17.16 &17.10 & \\
        17.58&17.80 &      &17.84 &17.94 & \\
        18.09&18.56 &      &      &18.47 & \\
        18.67&      &18.79 &18.66 &18.66 &18.71 \\
        19.29&      &19.22 &      &19.32 & \\
        19.87&      &19.69 &      &      & \\
             &19.88 &20.02 &20.03 &19.92 &19.94 \\
             &21.58 &21.28 &21.49 &21.46 & \\
             &      &21.80 &      &22.07 & \\
             &22.61 &22.37 &22.48 &22.58 &22.66 \\
             &      &      &22.75 &      & \\
        \noalign{\smallskip}
        \hline
        \noalign{\smallskip}
        \end{tabular}
    \end{table}

\begin{figure*}[]
        \centering
                \includegraphics[width=17cm]{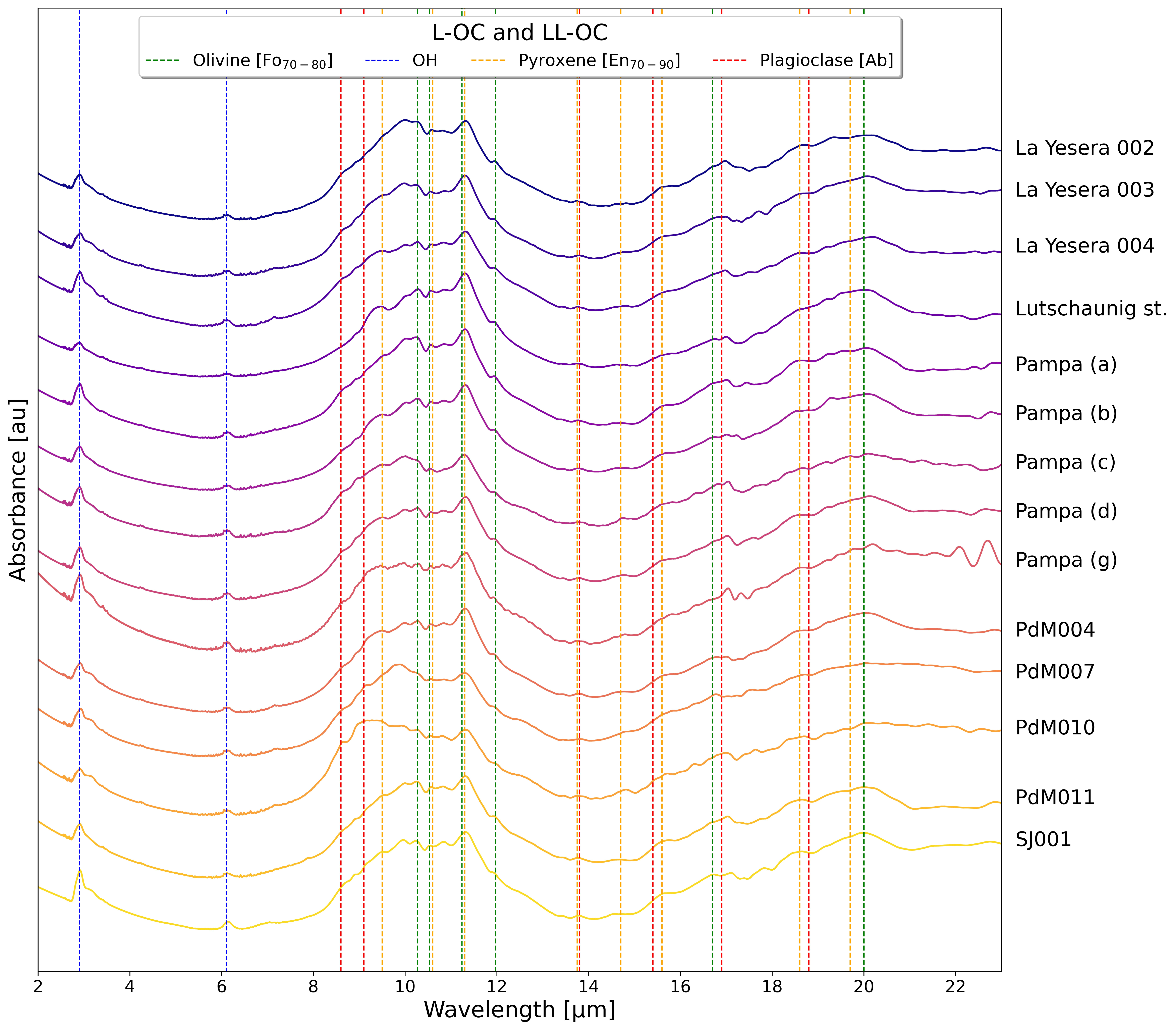}
          \caption{Spectra of ordinary chondrites: thirteen L class and one LL class (La Yesera 002). The spectra correspond to milled samples, with grain size between 1 - 10 \textmu m. The data presented are normalized to the maximum in the range of 8 - 12 \textmu m and vertically shifted for a better appreciation of shape and peaks. The vertical lines correspond to characteristic absorption bands of different minerals: light blue, for the water bands (OH); green for olivine bands, orange for the pyroxene bands, and red for plagioclase \citep{2002A&A...391..267C, 2003A&A...399.1101K, 2014Icar..229..263B, 2017P&SS..149...94C}.}
      \label{fig:abs-L}
\end{figure*}

\subsubsection{Absorbance of OC type L and LL} 
\vspace{0.3cm}
    The spectra of the 14 ordinary chondrites belonging to the L-OC and LL-OC groups are very similar to each other (see Fig. \ref{fig:abs-L}). The difference observed corresponds to variations in the relative intensity of certain absorption bands, except in a couple of samples that present considerable variations that do not appear in the other meteorites.
    One sample corresponds to the LL group, and the remaining 13 correspond to the L group. The spectra of L-OC and LL-OC are shown together for simplicity; they look very similar and have a similar composition. However, it is not possible to determine any further conclusions related to the difference between L-OC and LL-OC due to the limited number of LL samples.  
    The general spectra of L-OC and LL-OC maintain characteristics similar to those of the H-OC spectra, with shifts that could be assigned to chemical and compositional differences between the minerals. However, the L-OC and LL-OC spectra look significantly more diverse than within the CO3 and H-OC. Hence, it is important to consider a sampling bias as a result of the number of samples of each chondrite group studied in this work.

\begin{table*}
    \centering
    \caption{Peak positions (in \textmu m) for the absorption bands in our L-OC and LL-OC spectra.} 
    \label{tbl:abs-L}
    \scriptsize
        \begin{tabular}{lccccccccccccc}
        \noalign{\smallskip}
        \hline
        \noalign{\smallskip}      
        \textbf{PdM}&\textbf{La Yesera}&\textbf{PdM}&\textbf{La Yesera}&\textbf{La Yesera}&\textbf{PdM}&\textbf{PdM}&\textbf{Lutschaunig's}&\textbf{Pampa}&\textbf{Pampa}&\textbf{Pampa}&\textbf{Pampa}&\textbf{Pampa}&\textbf{San Juan}\\ 
        \textbf{004}&\textbf{002}&\textbf{007}&\textbf{003}&\textbf{004}&\textbf{010}&\textbf{011}&\textbf{stone}&\textbf{(a)}&\textbf{(b)}&\textbf{(c)}&\textbf{(d)}&\textbf{(g)}&\textbf{001}\\
        \noalign{\smallskip}
        \hline
        \noalign{\smallskip}
        2.52 &2.55 & 2.58& 2.55& 2.53& 2.60& 2.58& 2.52& 2.58& 2.55& 2.56& 2.58& 2.53& \\
        2.67 &2.67 & 2.71& 2.67& 2.67& 2.68& 2.65& 2.68& 2.69& 2.67& 2.68& 2.67& 2.68& \\
        2.90 &2.90 & 2.92& 2.90& 2.92& 2.90& 2.90& 2.90& 2.86& 2.91& 2.92& 2.91& 2.92& 2.91\\
             &    & 3.15& 3.08& & 3.08& & & 2.91& & 3.36& & 3.19& \\ 
             &3.42 & 3.41& 3.42& 3.42& 3.41& 3.41& 3.41& 3.41& 3.41& 3.41& 3.41& 3.42& 3.41\\ 
             & & & 3.50& 3.50& & & 3.50& 3.49& & & & 3.50& \\ 
             & & 4.23& 4.23& 4.23& 4.23& 4.23& & & & & 4.23& 4.24& \\ 
        4.27 &4.28 & 4.27& & & & & 4.27& 4.28& & 4.27& 4.27& & \\ 
             & & & 5.99& 6.07& 6.07& 6.07& 6.07& & & & & & \\ 
        6.11 &6.11 & 6.16& 6.16& & 6.16& 6.11& & 6.16& 6.11& 6.11& 6.11& 6.11& 6.12\\ 
        7.21 &7.16 & 7.22& 7.21& 7.16& 7.21& 7.62& 7.48& 7.35& 7.20& 7.47& 7.47& 7.47& 7.05\\ 
             & & & & & 8.65& & & & & & & 8.67& 8.94\\ 
             & & & & & 9.19& & & & & & 9.00& & \\ 
             & & & & & 9.46& & 9.48& & 9.50& 9.49& & 9.47& \\ 
        9.50 & & & & 9.53& & 9.56& & & & 9.57& 9.54& & 9.54\\ 
        9.99 &10.01& 9.90& 9.98& 10.01& 9.91& 10.00& & 10.04& 10.03& 10.01& 10.01& 9.98& 9.96\\ 
        10.28&10.25& & 10.25& 10.28& 10.27& 10.26& 10.28& 10.26& 10.28& & 10.27& 10.31& 10.25\\ 
        10.56&10.59& 10.56& 10.56& 10.58& 10.55& 10.58& 10.57& 10.57& 10.57& 10.55& 10.56& 10.56& 10.55\\
        10.85&10.82& 10.86& 10.86& 10.85& 10.82& 10.84& 10.85& 10.85& 10.86& & 10.85& 10.82& 10.85\\ 
        11.32&11.33& 11.32& 11.32& 11.32& 11.34& 11.32& 11.31& 11.32& 11.32& 11.29& 11.32& 11.33& 11.32\\ 
        11.92&11.95& & 11.93& 11.93& & 11.94& 11.94& 11.94& 11.93& & 11.94& 11.93& \\ 
        13.44&13.46& & 13.43& 13.46& 13.44& & & 13.46& & 13.45& & 13.43\\ 
        13.79&13.75& 13.83& 13.80& 13.76& 13.76& 13.78& 13.73& 13.78& 13.78& 13.78& 13.79& 13.77& 13.80\\ 
             & & & & & 14.17& & & & & & & & \\ 
             &14.61& & & & & & 14.60& 14.57& & & & & 14.57\\ 
        14.78&14.83& & 14.82& 14.82& 14.82& 14.71& 14.77& 14.73& 14.67& 14.76& 14.79& 14.81& \\
        15.90&15.83& 15.87& & & 15.73& 15.77& 15.82& & & 15.82& 15.83& 15.84& 15.84\\ 
             & & & & & 16.32& & & & & 16.38& & & \\ 
        16.79&16.98& 16.79& 16.87& 16.98& 16.99& & 16.76& & 16.76& 16.84& & & 16.74\\ 
        17.01& & 17.06& 17.00& & & 17.03& 17.00& 17.01& 17.02& 17.05& 17.04& 17.05& 17.10\\ 
        17.30& & 17.24& 17.23& & & & & & 17.22& & & & \\ 
             & & 17.59& 17.50& 17.52& 17.65& & & 17.46& & & 17.60& 17.67& 17.40\\ 
             & & 17.84& 17.72& & & 17.79& & & & 17.83& & & 17.84\\ 
        18.73&18.72& 18.73& 18.73& 18.62& 18.72& 18.68& & 18.64& 18.69& 18.70& 18.73& 18.74& 18.66\\
             & & & 18.82& & & & & & & & & & \\ 
             &19.37& & & 19.23& 19.24& & 19.22& 19.29& 19.30& 19.21& 19.23& 19.23& \\ 
             & & & & 19.26& 19.27& & & & & & & 19.48& \\ 
             & & & & & & & & 19.60& & 19.72& & 19.83& \\ 
        20.07&20.16& 20.05& 20.09& 20.23& 20.10& 20.04& 20.08& 20.07& 20.10& 20.12& 20.16& 20.21& 20.00\\ 
             & & & & & 20.50& & & & & & & & \\ 
             & & 20.77& & & 20.79& & & & & 20.81& & 20.73& \\ 
             & & 21.15& & & & & & & & 21.27& & 21.16& \\
        21.48& & 21.49& & 21.53& 21.42& & 21.59& 21.52& & & 21.41& 21.54& 21.57\\ 
        21.95&21.71& 21.89& 21.65& 22.44& 21.97& 21.75& 22.06& 22.44& 21.75& 21.76& 21.72& 22.09& 22.05\\ 
             & & & 22.35& & & & & & 22.23& 22.25& & & \\ 
        22.86&22.66& & & & 22.56& 22.87& 22.85& 22.87& 22.78& & 22.68& 22.71& 22.77\\ 
        \noalign{\smallskip}
        \hline
        \end{tabular}
    \end{table*}

\begin{figure}[]
        \centering
           \includegraphics[width=8cm]{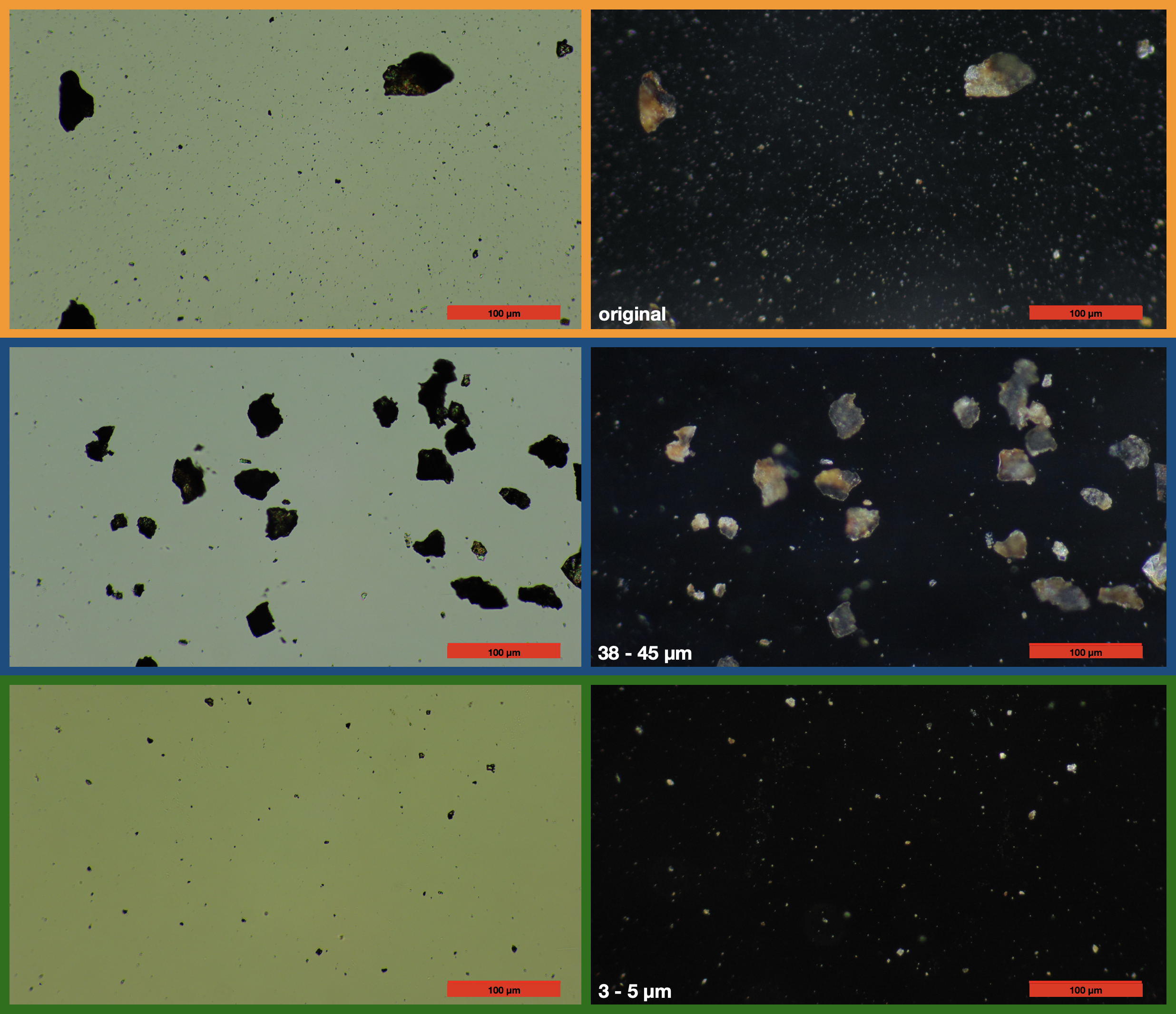}
          \caption{Microphotography of sample Los Vientos 123 with three different grain size distributions at 10x magnification. The left side is bright field mode (TLBF) and the right side is dark field mode (TLDF). The first row corresponds to the original sample; the second row is the sieved sample with particles in the range 38 - 45 \textmu m; the last row is the sample milled 60 - 75 minutes with particles smaller than 10 \textmu m. It is possible to observe a large fraction of small particles in the original sample. These small particles dominate the spectra compared to the larger particles. Therefore, the spectrum of the original sample is very similar to the spectrum of a milled sample with only small particles <10 \textmu m.}
      \label{fig:los-vientos-123-microphot}
\end{figure}

\begin{figure}[]
        \centering
                \includegraphics[width=9cm]{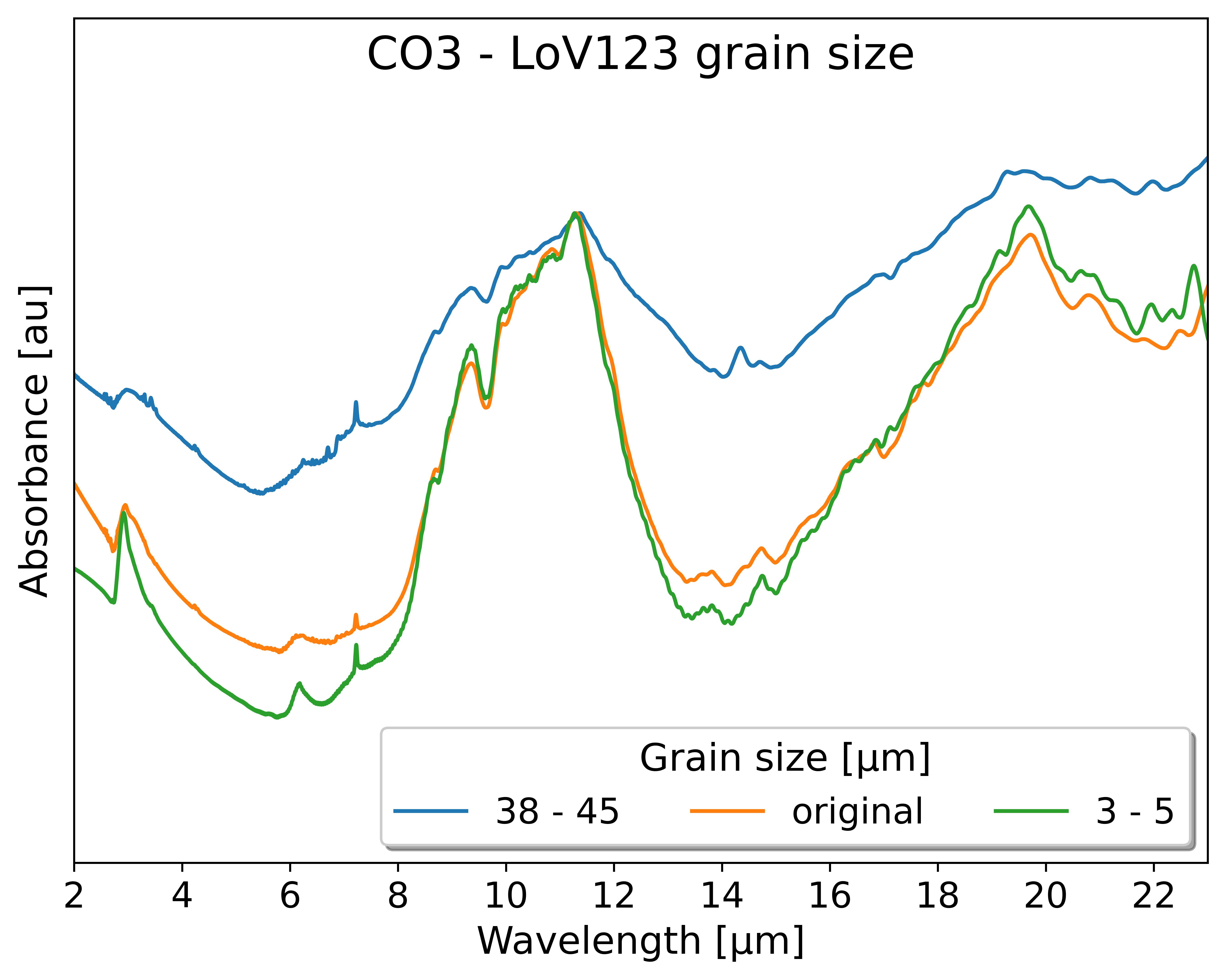}
          \caption{Absorbance spectra of three samples of Los Vientos 123 meteorite with three different grain size distributions. The blue line corresponds to the sieved samples 38 to 45 \textmu m particles, the red line corresponds to the original samples, and the black line corresponds to the dust of 1 to 10 \textmu m (grounded for 60-75 minutes) in the agate mortar.}
      \label{fig:los-vientos-abs-size}
\end{figure}

\begin{figure*}[]
        \centering
                \includegraphics[width=8.9cm]{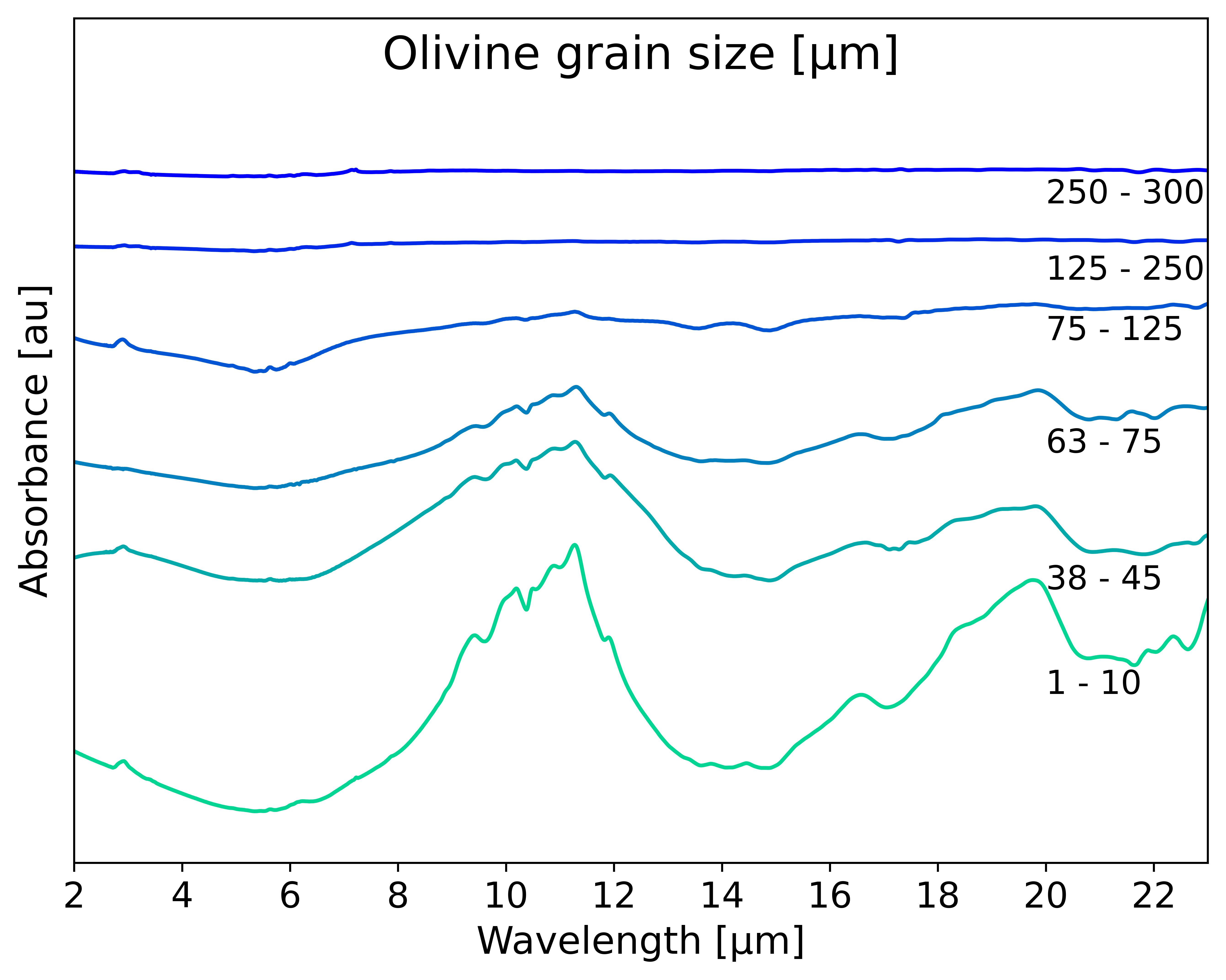}
        \includegraphics[width=8.9cm]{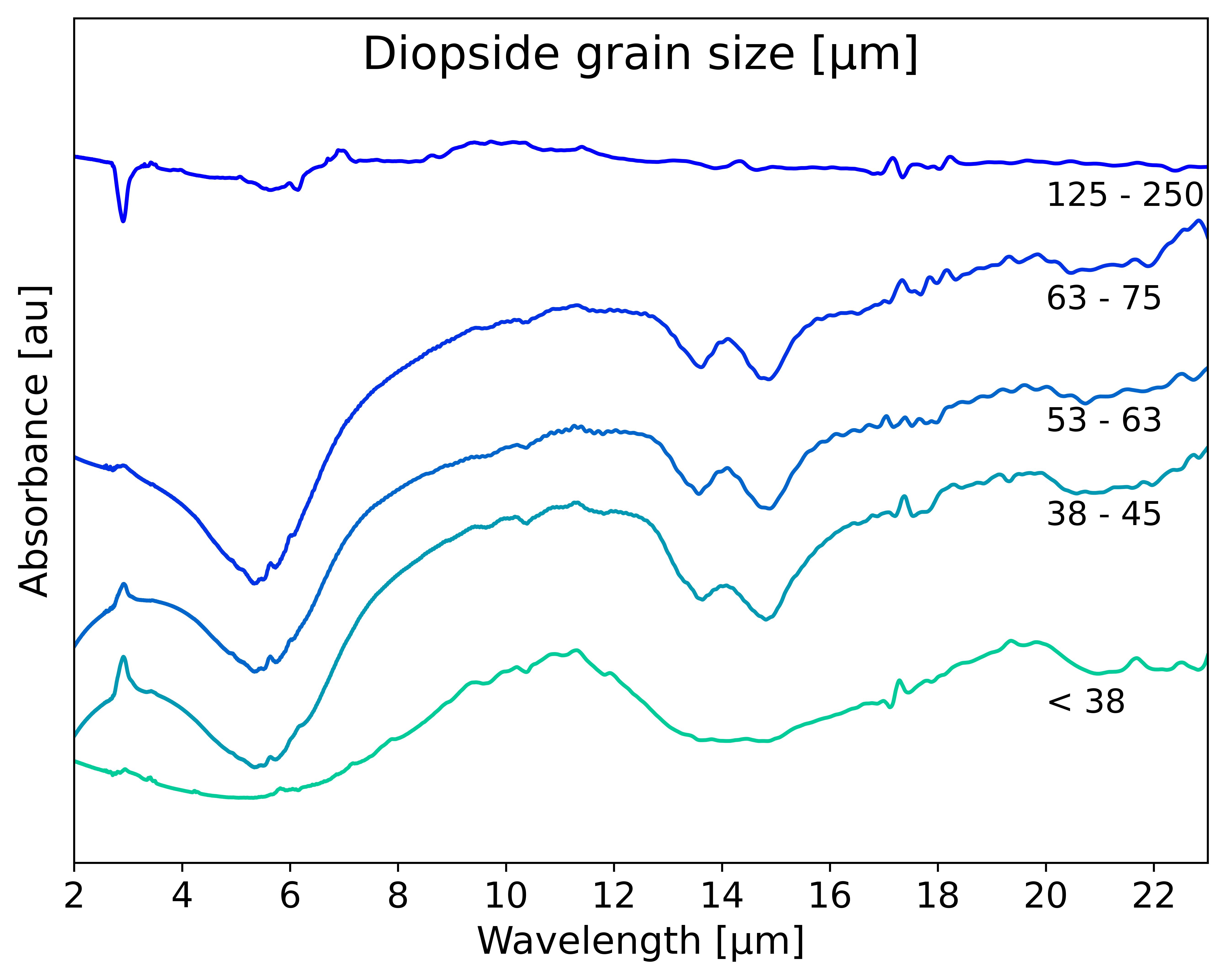}
          \caption{Absorbance spectra of pure minerals with different grain size distributions. Range of size distribution: G-H) 300 – 250 \textmu m, F-G) 250 – 125 \textmu m, E-F) 125 – 75 \textmu m, D-E) 75 – 63 \textmu m, A-B) 45 – 38 \textmu m, A) <38 \textmu m. Spectra are presented vertically shifted for comparison.}
      \label{fig:pure-abs-size}
\end{figure*}

    The L-OC and LL-OC spectra show a complex shape of several peaks around the 10 \textmu m area, corresponding to the silicate mixture mainly dominated by a well-defined sharp peak at 11.3 \textmu m, belonging to olivine, along with peaks at 10.2, 11.9 and 20.1 \textmu m \citep{2003A&A...399.1101K, 2007PCM....34..319H, 2010MNRAS.406..460P}.
    Compared to H-OC, L-OC, and LL-OC show a greater difference in the relative intensity of the peaks between 10.2 and 11.2 \textmu m, and the shifts of the olivine absorption bands, which would indicate the greater chemical variation between Fe-Mg of this mineral, although they remain to be Mg-rich olivine than Fe-rich olivine. 

    Around 20 \textmu m, the absorption band is shifted between 19.9 and 20.1 \textmu m as a product of the variation of the Fe content in olivine. In some spectra (SJ001, Pampa (b), Pampa (a), Lutschaunig stone, PdM011), the peak becomes narrower and more pointed, similar to the observed in H-OC, while in other curves (Pampa (g), PdM010, PdM007) this feature is wider and less defined, which is also consistent with the amount of forsteritic olivine \citep[Mg;][]{2003A&A...399.1101K, 2007PCM....34..319H, 2010MNRAS.406..460P}.
    Likewise, the small absorption bands between 13 and 15 \textmu m can be associated with pyroxene enstatite along with peaks at 9.5, 10.3, 10.7, 15.7, and 18.6 \textmu m. 

    Slight shifts and intensity variations between samples could be associated with three main factors: i) crystallographic difference between ortho-enstatite (orthorhombic) and clinoenstatite (monoclinic), ii) compositional variations between Mg-rich enstatite and Fe-rich ferrosilite solid solution \citep{2000A&A...363.1115K, 2002A&A...391..267C, 2007MNRAS.376.1367B}, and iii) variations in the abundance of the different mineral phases.
    Some samples, Pampa (g), PdM010, and PdM007 present less defined peaks ("bumps") around 9 \textmu m and flatter signatures around 10 \textmu m. These samples also present a relatively featureless region around 20 \textmu m. The bands at 10.42 and 19.23 \textmu m are characteristic of amorphous enstatite \citep[glass;][]{2000A&A...363.1115K}, which are accentuated in these spectra. This probably corresponds to a high contribution of noncrystalline material (amorphous silicates or glass) with a composition of silica, olivine, and pyroxene.

    In some samples, it is possible to observe a peak at 8.7 \textmu m, indicating the presence of plagioclase feldspar, along with bands at 9.9, 10.9, 16.0, 17.2, and 18.5 \textmu m. Variations in the position of these peaks correspond to variations in composition within the albite–anorthite solid solution \citep{2017P&SS..149...94C}.
    The well-defined feature at 9.97 \textmu m that appears in H-OC is also observable in some of these samples. In particular, Pampa (c) and La Yesera 002, show a stronger band compared to other L-OC  and LL-OC.

    Considering that all our samples were collected from the same areas, it is possible that some are paired samples. This could affect the diversity of spectra, as paired samples are more likely to have similar spectra. For instance, based on similar classifications, spectra, and terrestrial ages, Pampa (c), Pampa (d), and Pampa (g) might be paired \citep{2017M&PS...52.1843P}.

    Several absorption bands can be associated with more than one mineral, which is repeated in L-OC and LL-OC. The details regarding the absorbance peaks of each sample are presented in Table \ref{tbl:abs-L}.

\vspace{0.3cm}
\subsection{Absorbance and grain size}\label{absgrainsize} 
\vspace{0.3cm}
    All the absorbance spectra shown in the previous sections correspond to the original dust samples, which contain a range of particle sizes.  In order to characterize their size distribution, we used the microphotography technique described in Sect. \ref{Micro-photography}. We find that the original samples contain mostly micron-sized grains, but some samples present a few grains that are larger, up to $\sim$100 \textmu m (see Fig. \ref{fig:los-vientos-123-microphot}, top panel). In order to produce two different size distributions, we used the pair of sieves A-B to collect particles between 38 and 45 \textmu m. We also ground the small grains for 60 - 75 minutes in an agate mortar.  

    Using the two distinct samples (sieved and ground grains), we created new KBr pellets and performed additional absorbance measurements. In Fig. \ref{fig:los-vientos-abs-size}, we compare the spectra of the sieved and ground grains with the original spectra. We find that the spectrum of the ground grains remains very similar to that of the original sample, while the sieved sample containing only large grains shows much weaker features. 
    After dissolving the new pellets, we obtained microphotography images corresponding to sieved particles and ground particles, which are also shown in Fig. \ref{fig:los-vientos-123-microphot}, middle and bottom panels, respectively.
    We find that the size distributions of both the original and ground samples have a similar peak, which explains the very similar absorbance spectra. This means that the small grains dominate the opacity even though most of the mass might be in a few much larger grains.

\subsection{Absorbance and grain size in pure minerals}
\vspace{0.3cm}
    The absorbance spectra of pure minerals as a function of grain size are shown in Fig. \ref{fig:pure-abs-size}. As seen with meteorites, the main effect of increasing the grain size is a reduction in the strength of the features, some of which might even disappear.

\subsection{Opacities}
\begin{figure}[]
        \centering
                \includegraphics[width=9cm]{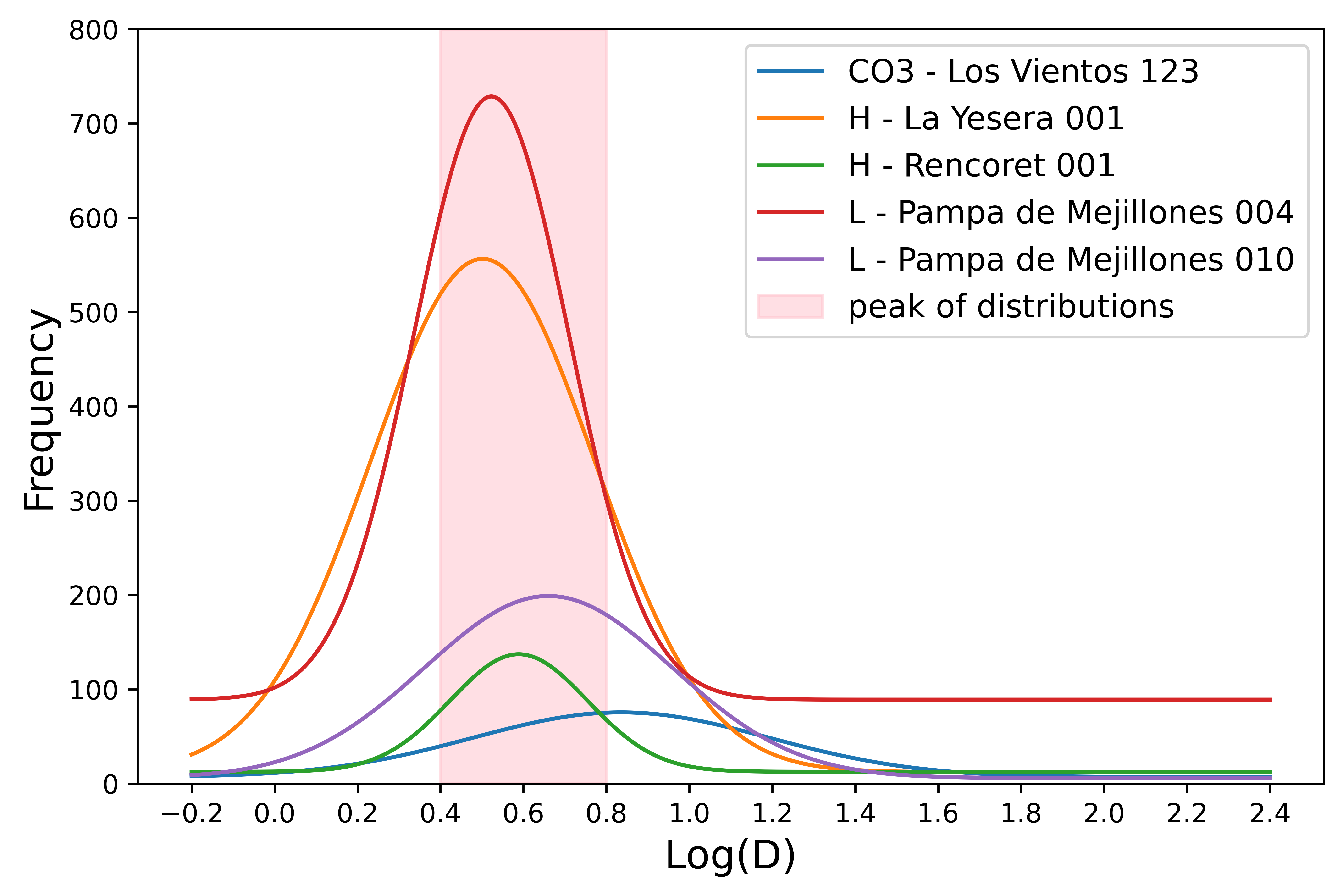}
          \caption{Grain size distribution of CO3s, H-OC, L-OC, and LL-OC grounded for 60 - 75 minutes in an agate mortar. The red area is center in the peak of the distributions, and the grain size is between 2.5 and 5.5 \textmu m.}
      \label{fig:all-meteorites-sizes}
\end{figure}

\vspace{0.3cm}
    Since the opacity at MIR wavelengths is dominated by small particles (a few microns in size), and given the limited total mass of meteorite samples available in the laboratory, we focused on measuring the opacities of finely ground grains. To achieve this, all samples were ground for 60 - 75 minutes in an agate mortar, resulting in particle size distributions peaking at 3 to 5 \textmu m for each meteorite type (see Fig. \ref{fig:all-meteorites-sizes}).

\begin{figure*}[]
        \centering
                \includegraphics[width=15cm]{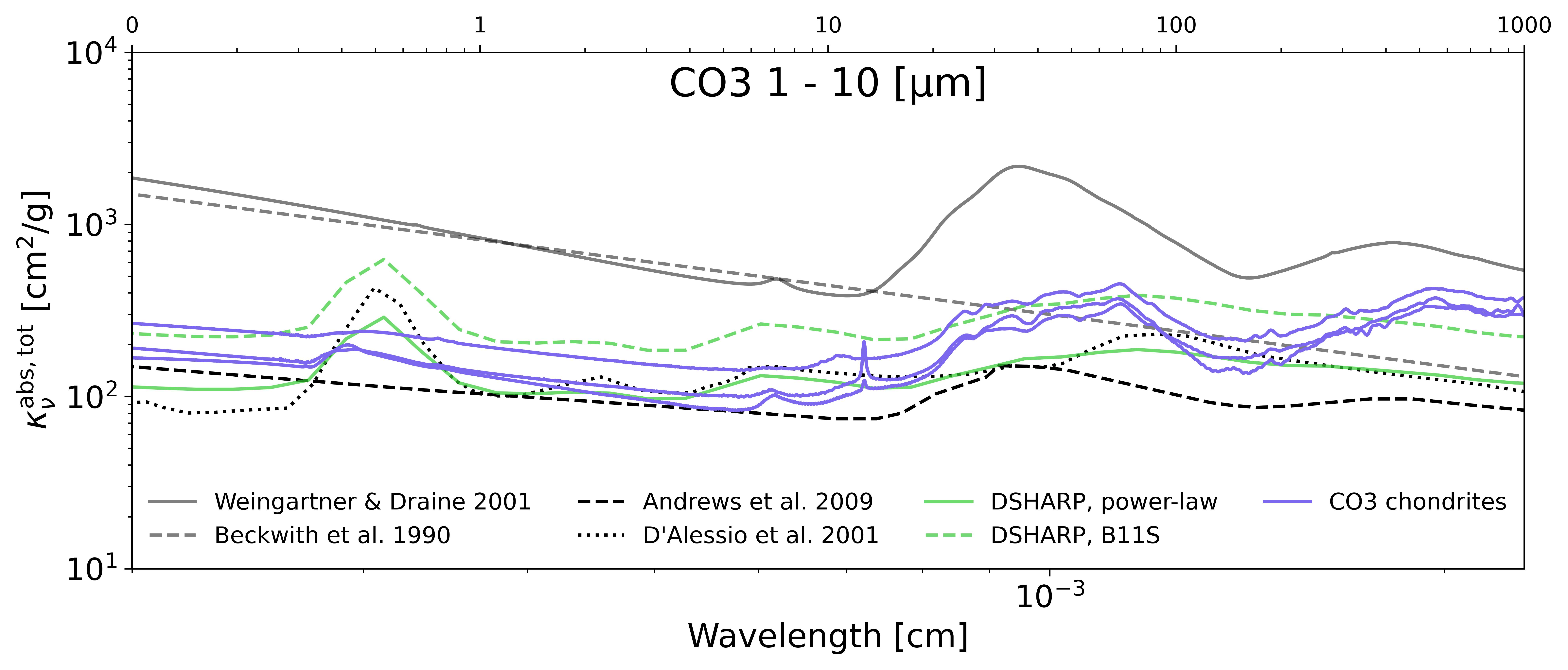}
        \includegraphics[width=15cm]{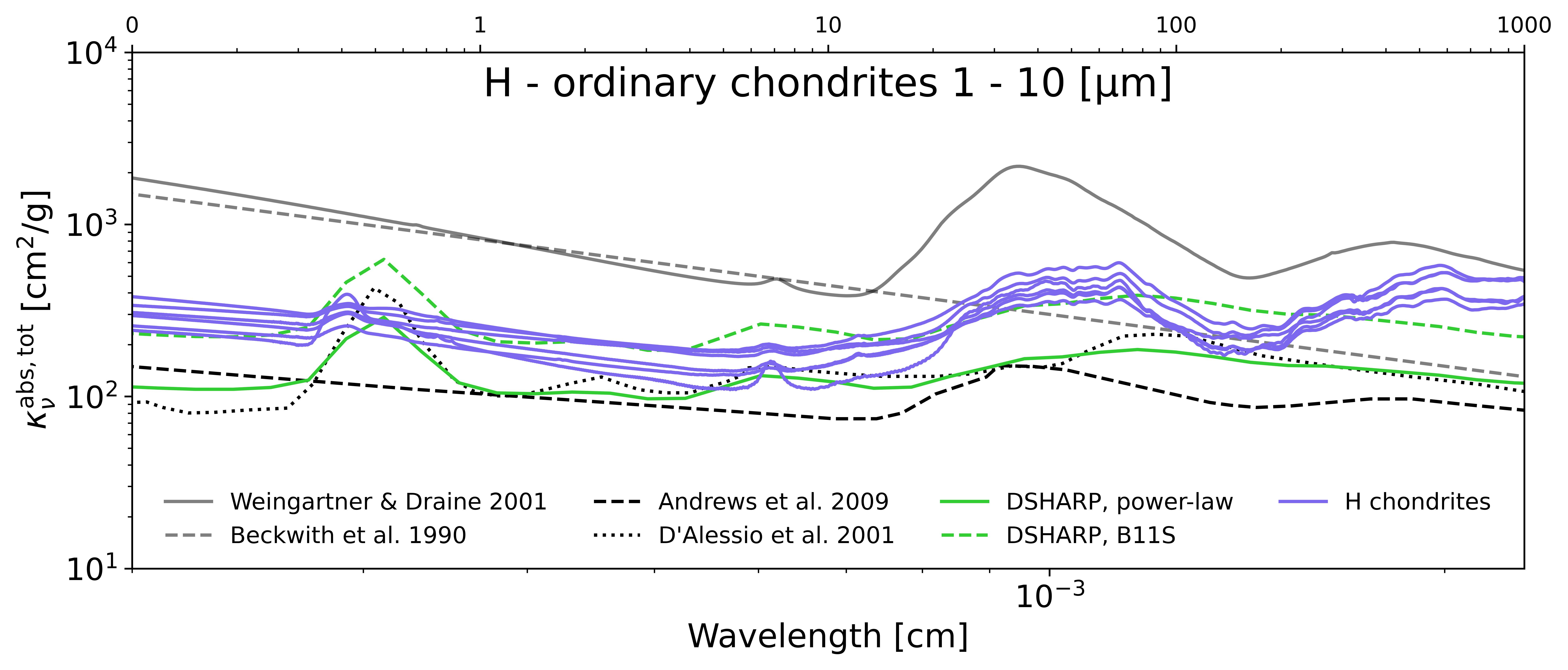}
        \includegraphics[width=15cm]{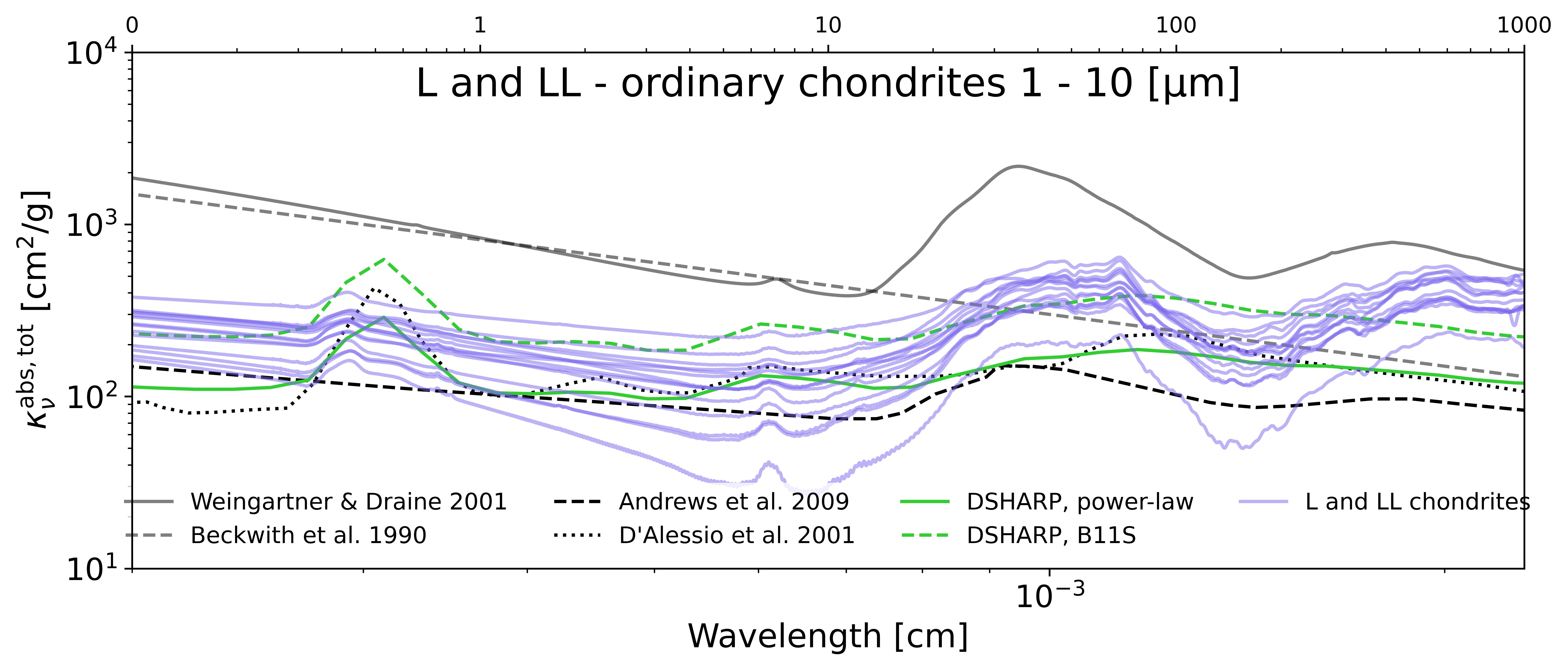}
          \caption{Comparison of opacities from this work of CO3s meteorites (upper), H ordinary chondrites (middle) and L and LL ordinary chondrites (bottom) with grain size between 1 - 10 \textmu m to opacities from the literature \citep{2018ApJ...869L..41A}}.
      \label{fig:DSHARP-met}
\end{figure*}

    The resulting opacities from the grounded meteorites are shown in Fig. \ref{fig:DSHARP-met} for CO3s chondrites, H-OC, L-OC, and LL-OC. They are compared with values from the literature compiled by the DSHARP program \citep{2018ApJ...869L..41A, 2018ApJ...869L..45B}. The most relevant comparison corresponds to the values provided by \cite{2001ApJ...548..296W}, which correspond to small grains with a$_{max}$ $<$ 0.1 \textmu m of the interstellar medium (ISM). 
    The fact that our grain size distribution extends to a few microns explains why our opacities are lower than ISM values, and they are likely to be more representative of the grain size distribution observed in protoplanetary disks.

\begin{figure*}[]
        \centering
                \includegraphics[width=15cm]{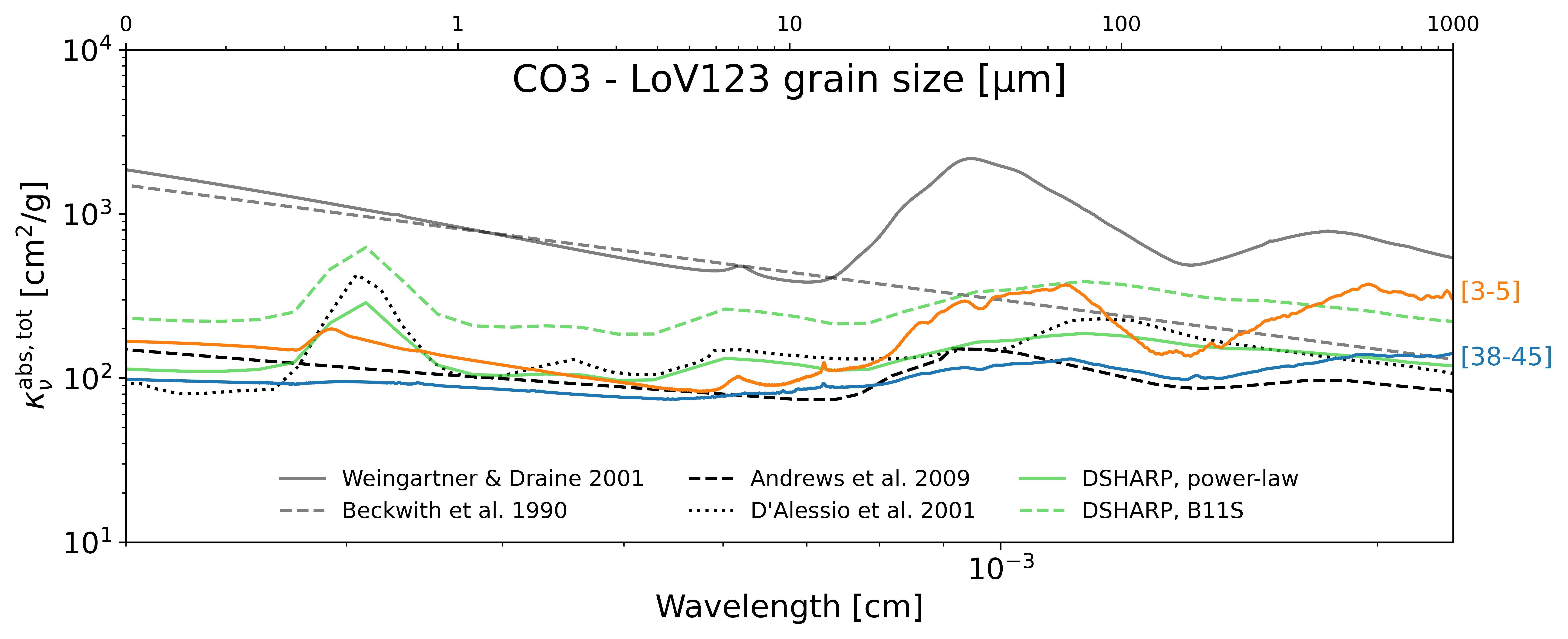}
        \includegraphics[width=15.5cm]{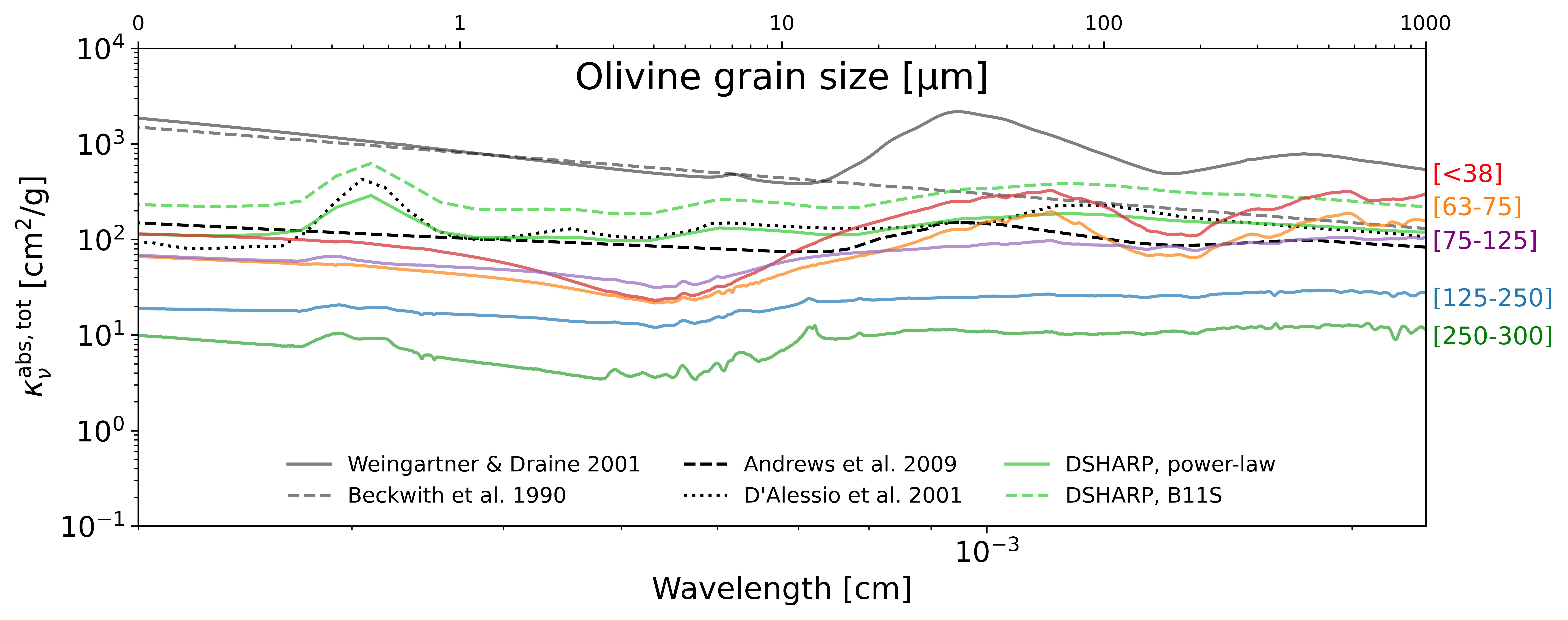}
          \caption{Comparison of Los Vientos 123 opacities (upper) and opacities from the literature. We show two different ranges of particle size: the blue line indicates a sieved sample with a range of 38 to 45 \textmu m, and the orange line indicates the grounded sample with a range of 1 to 10 \textmu m, with peak grain size between 3 - 5 \textmu m. Comparison of olivine opacities (bottom) to opacities from the literature. We show different range sizes from 1 - 10 \textmu m to 250 - 300 \textmu m.}
      \label{fig:DSHARP-size}
\end{figure*}

    For the CO meteorite Los Vientos 123, we had enough material to measure opacities with grain sizes between 38 and 45 \textmu m. As expected, we find that the larger grains have a significantly lower opacity than micron-sized grains (see Fig. \ref{fig:DSHARP-size}, upper panel). 
    We also performed a similar analysis using pure olivine, extending the size distributions to grains as large as 300 \textmu m, at which points the MIR opacity decreases dramatically (see Fig. \ref{fig:DSHARP-size}, bottom panel).

    The values of opacities and absorbance spectra available for all samples correspond to grain sizes smaller than 10 \textmu m, with a size peak between 3 and 5 \textmu m. In addition, Table \ref{tbl:opacities-available} lists the samples for which opacities and absorbance spectra are available for different grain size distributions. The data set is in the repository of the UDP Cosmic Dust Laboratory.

\begin{table*}
    \centering
    \caption{Meteorites and minerals with absorbance spectra and opacities from the UDP Cosmic Dust Laboratory. These data will be available through the UDP repository, \href{https://astronomia.udp.cl/en/collaboration/udp-cosmic-dust-laboratory/}{Cosmic Dust Laboratory}.}
    \label{tbl:opacities-available}
        \begin{tabular}{lccccccccc}
        \noalign{\smallskip}
        \hline
        \noalign{\smallskip}
        \textbf{Sample}&\textbf{38-45}&\textbf{53-63}&\textbf{63-75}&\textbf{75-125}&\textbf{125-250}&\textbf{250-300}\\
        &\textbf{[\textmu m]}&\textbf{[\textmu m]}&\textbf{[\textmu m]}&\textbf{[\textmu m]}&\textbf{[\textmu m]}&\textbf{[\textmu m]}\\ 
        \noalign{\smallskip}
        \hline
        \noalign{\smallskip}
        Olivine   & $\star$ &         & $\star$ & $\star$ & $\star$ & $\star$ \\
        Diopside  & $\star$ & $\star$ & $\star$ &         & $\star$ \\
        \noalign{\smallskip}
        \hline
        \noalign{\smallskip}
        Los Vientos 123 & $\star$ \\
                
        \noalign{\smallskip}
        \hline
        \end{tabular}
    \end{table*}

\section{Discussion}\label{section_4}
\vspace{0.3cm}
    The features of absorbance spectra that appear in all samples are discussed in this section. We briefly mention the relations between the mineralogy of chondrites with the petrological grades, shock stages, and weathering grades of meteorites. However, a more detailed analysis is beyond the scope of this work, and we do not analyze patterns related to these relations with the spectra of our samples.

\subsection{Absorbance of CO3s chondrites}
\vspace{0.3cm}
    The CO3 meteorites exhibit distinct absorption bands around 3.0 \textmu m and 3.4 \textmu m. These bands correspond to the stretching vibrations of OH and CH bonds, respectively, indicating the presence of water molecules and organic compounds \citep{2002ApJ...566L.113G}.
    Additionally, the 3.0 \textmu m band, along with features observed around 6 - 7 \textmu m, may be associated with terrestrial weathering phases, such as sulfates or carbonate minerals \citep{2003asdu.confE..72K}.

    The CO3s present peaks that correspond to carbon-related molecules, which are not found in our other chondritic meteorites. In particular, in sample Catalina 008 (C008), the peak at 7.25 \textmu m is well-defined, very sharp, and stronger than seen in the other two samples. The intensity of this feature in C008 could be due to the larger amount of carbon material in this sample compared to EM216 and LoV123.
    Organics tend to be richer in the matrix than in the chondrules. This stronger peak of carbon material in C008 could be due to the higher abundance of matrix in C008 than in the other two CO3s. The bulk matrix abundance for each meteorite reported in The Meteoritical Bulletin is 58\% vol of the matrix in C008, 36\% vol of the matrix in LoV123, and EM 216 has a chondrule: matrix ratio of around 1: 1, which implies that the matrix abundance is approximately $\leq$ 50\%.

    However, it is not clear what compound is responsible for this strong peak. We discuss some options related to observations of similar features in different objects.

    Different carbonate minerals (calcite, aragonite, dolomite, magnesite, siderite, and rhodochrosite) show a strong feature around 7 \textmu m, but this band appears to be broader than the peak of 7.25 \textmu m presented in our CO3s samples \citep{2003asdu.confE..72K}.  
    In \cite{2012ApJ...747...93L}, they show similar features of carbonaceous material observed with  \emph{Spitzer} in the disk around $\eta$ Corvi that could be attributed to nano-diamonds, according to measurements performed by \cite{2010DRM....19.1207G}, showing spectra dominated by the main diamond C–C stretch at 7.51 \textmu m, by ppm C–N impurity vibrational features at 6.99 \textmu m, by platelets at 7.34 \textmu m, and by C–H stretches at 7.11, 3.59, 3.21, 3.09, and 2.22 \textmu m.

    The 7.25 \textmu m peak can also be attributed to other organic molecules. For instance, HCOOH has a characteristic peak at 7.25 \textmu m due to the C-H bond \citep{1999A&A...343..966S}. Similarly, H$_{2}$CO has a peak at 5.81 \textmu m and shows a peak at 6.68 \textmu m, a small peak that can be observed in EM216. 
    According to \cite{2004A&A...416..165R}, astronomical observations of the 7.25 \textmu m peak are assigned to OCN-, which is usually seen at 4.62 \textmu m. They proposed that both must be due to the presence of HNCO. However, 4.62 \textmu m is not a peak present in our meteorite samples.
    Alternatively, there is the 5.83 \textmu m peak which is usually related to ice, but \cite{2001A&A...376..254K} mentioned that the intensity of this peak could not be fully explained by ice, since with other ice peaks there is a 40\% excess of intensity only at that peak (which is not present in other peaks of H$_{2}$O ice), and this could be explained by the presence of organic matter containing the C-O group.
    According to \citep{2001A&A...376..254K}, the peak at 5.83 \textmu m is characteristic of the vibration mode of the C-O bond of the carbonyl group (ketones, aldehydes, carboxylic acid, and esters). This peak is slightly apparent in LoV123 but not in the other samples. However, sometimes samples that do not have a peak at 5.8 \textmu m do present a strong peak near 7.24 \textmu m, which is observed in all CO3s chondrite samples \citep{2001A&A...376..254K}. 
    All of our meteorite samples are referred to as "finds." In hot deserts such as the Atacama, sulfates, oxyhydroxides, Ca carbonates, and silica can be formed by weathering \citep{2006mess.book..853B}. This terrestrial weathering product can contribute particularly to the 3 \textmu m region.

    The features around 10 and 20 \textmu m are assigned to a mixture of silicates, with a major content of forsterite-fayalite \citep{2003A&A...399.1101K, 2007PCM....34..319H, 2010MNRAS.406..460P}, the presence of pyroxene of intermediate composition En$_{50}$ \citep{2000A&A...363.1115K, 2002A&A...391..267C, 2007MNRAS.376.1367B} and less abundant albite-anorthite plagioclase feldspar \citep{2017P&SS..149...94C}. The predominance of olivine implies that it is found in greater relative quantities than the other minerals.
    It is possible to observe a difference between the peaks in the 8 - 9 \textmu m range. This peak may be due to the influence of different plagioclase feldspars: In LoV123 a single peak is observed at 8.7 \textmu m, which may indicate a greater influence of plagioclase anorthite, while in EM216 and C008, two peaks are observed at 8.7 and 8.9 \textmu m, which would correspond to plagioclase with albitic composition \citep{2017P&SS..149...94C}.

    In the case of LoV123, between 10 and 11 \textmu m, several peaks are present that are not as pronounced in the other two CO3 chondrites. This suggests that LoV123 contains a higher proportion of the silicate mixture, leading to more prominent peaks in this range. As a result, LoV123 shows a stronger influence of pyroxene compared to the other two CO3 chondrites.

\subsection{Absorbance of OC}
\vspace{0.3cm}
    Ordinary chondrites present similar mineralogical characteristics at the group level, and their differentiation is at a more specific level. Although our ordinary chondrites H, L, and LL are spectrally similar, \cite{1990GeCoA..54.1217R} noted differences regarding olivine composition: molar percentage of Fayalite (\%mol Fa); and the relative abundance of Co in kamacite vs Fayalite in olivine. In addition, H, L, and LL chondrites differ in their degree of oxidation, with H being the most reduced and LL the most oxidized. Metals do not absorb clearly in the MIR range, which may contribute to the similarities of H, L, and LL spectra \citep{1968PhRv..166..667B, stuart2004infrared, dombrovsky2011radiative, hummel2011electronic,  2017AdPhX...2..373Y}.

    Groups are valuable for distinguishing meteorites of the same class, as they reflect variations in oxidation state during their formation. Differences in oxidation state suggest that meteorites with different petrological types formed in distinct chemical environments within the solar nebula, shaped by factors such as the local availability of oxygen, temperature, and proximity to the Sun \citep{1990GeCoA..54.1217R}. It is important to note that Fe oxidation can occur in asteroidal environments, but not within the parent body of ordinary chondrites of L 5/6 \citep{2006mess.book...19W}. Abundances of siderophile elements are also used to distinguish from ordinary chondrites: while the H types are the chondrites with the highest content of these elements, LL are the ordinary chondrites with lower amounts of these elements \citep{1989GeCoA..53.2747K, 1991LPSC...21..493S}.
    This explains why the observed spectra of H-OC, L-OC, and LL-OC generally exhibit a similar overall shape, characterized by peaks and bands corresponding to the same major minerals (olivine, pyroxene, and plagioclase), although with minor compositional variations. Table \ref{tbl:XRD} provides the measured compositions of thirteen samples, emphasizing compositional similarities among certain meteorites. For instance, La Yesera 004 and PdM 004, classified as L-OC, display nearly identical compositions and closely matching spectra, as shown in Fig. \ref{fig:abs-L}. Similarly, Rencoret 001 and PdM 002 exhibit comparable trends within the H-OC group.

    \cite{1990GeCoA..54.1217R} infers that chondrite types 3 - 4 accreted less FeO than chondrites type 2 or fine matrix material with a lower FeO content.
    The shift in the peaks is directly correlated with the amount of Fe present and the precise composition of the silicates. Mostly olivine and pyroxene of intermediate composition occur, which does not exclude the fact that minerals with composition from the extremes of both solid solutions may be present in smaller quantities and are masked by the predominant intermediate composition in the sample.

    The spectra of Pampa (g) and PdM010 show broad peaks that are characteristic of amorphous silicates, which typically exhibit a featureless spectrum in the infrared range \citep{1995A&A...300..503D, 2005SSRv..119....3M, 2017A&A...600A.123D, 2017A&A...606A..50D}. In particular, these two meteorite spectra look very similar to the spectra of the debris disk BD + 20307 \citep{2005Natur.436..363S}, where pristine glass - typically expected in meteorites with petrological type 3.00 - has been proposed as a possible component. Primitive amorphous phases, in small amounts, have also been reported in Semarkona (LL3.0) by \cite{2020M&PS...55..649D}. 
    However, primary amorphous silicates would have been of low abundance and likely restricted to the matrix, which constitutes only 15\% of OCs  \citep{1983JGR....88.9513Y, 1989GeCoA..53.2747K, 2005GeCoA..69.4907R, 2010GeCoA..74.4807R}. Additionally, primary amorphous material quickly recrystallizes with heating, which explains its absence in equilibrated OCs.

    The amorphous signatures observed in protoplanetary disks may not necessarily be strictly primary in origin. These glassy minerals could also form as a result of samples being subjected to varying shock pressures during collisions \citep{1991GeCoA..55.3845S}. 
    In the case of our samples, we interpret these amorphous phases as the result of deformations caused by shock events. 
    \cite{2010Icar..207...45M} conducted laboratory measurements on artificially shocked Murchison chondrite (CM), applying pressures ranging from 0 to 49 GPa. Their results demonstrated that glassy material can form from hydrated minerals through shock events, with recrystallization occurring over this glassy material at higher pressures. From our samples, Pampa (g) and PdM010 could be associated with samples subjected to pressures between 30 and 36 GPa, as they exhibit similar spectral features.
    However, CMs are richer in hydrated minerals, such as phyllosilicates, compared to OCs \citep{1985ClMin..20..415B, potin2020mineralogy}, which are predominantly composed of olivine and pyroxene, minerals that are significantly more resistant to deformation than phyllosilicates. Phyllosilicates are structurally weaker and more prone to deformation than olivine and pyroxene \citep{wang2001effective, 2003GeoJI.155..319V, mondol2008elastic}, with their transformation into amorphous phases primarily attributed to dehydroxylation, a process driven by the loss of structural water under shock conditions \citep{2005JMPeS.100..260N}. On the other hand, the deformation of olivine and pyroxene in OCs is likely mechanical or the result of shock melting during high-velocity collisions \citep{1991GeCoA..55.3845S}. Consequently, the shock pressure thresholds established by \cite{2010Icar..207...45M} may not fully apply to our OCs, given their distinct mineralogical composition and resistance to deformation. In our spectra, we conclude that many of the samples exhibit amorphous phases, which likely correspond to a multitude of low-crystallinity phases or minerals probably produced by shock processes.

    Phases of minor minerals, such as magnetite, sulfides, and other oxides, are not observed in the MIR absorbance spectra. This is due to the range in which our measurement was carried out, as oxides and sulfides typically do not show prominent features at these wavelengths \citep{2002A&A...393L.103H}. However, some oxide minerals, such as magnetite, exhibit features in the infrared range, including a band near 17.3 \textmu m attributed to Fe-O stretching vibrations \citep{jubb2010vibrational, li2012infrared}. These features are generally weaker than those of silicates, which dominate the MIR spectra due to their strong Si-O vibrational modes. For instance, \cite{2014Icar..229..263B} did not report magnetite-related bands in the KBr spectra of magnetite-rich CCs, such as Grosvenor mountains 95577, Bells, Essebi, Wisconsin Range 91600, and Niger. Consequently, the absence of detectable oxide and sulfide features in our spectra is consistent with their relatively low absorbance intensity in the MIR range.

\subsection{Absorbance and particle size}
\vspace{0.3cm}
    Aside from composition, the particle size distribution of the sample has an important effect on the shape of the MIR spectra. This can be observed in both pure minerals and meteorites.

    Figure \ref{fig:pure-abs-size} shows olivine and diopside spectra, respectively, with different ranges of size distributions. Spectra of large particle sizes (250 - 300 \textmu m for olivine and 125 - 250 \textmu m for diopside) are noisier and/or show less distinguishable features. This is because very little radiation is being absorbed in the KBr pellets that host the largest sample grain sizes, implying that we are predominantly measuring the KBr powder. The characteristic absorption features of olivine begin to appear from sizes less than 75 \textmu m. Spectra of smaller particle sizes (<10 \textmu m) are very well-defined and well-featured with sharper peaks.  
    In both cases, in the MIR measurement range, as the size of particles increases, the characteristic bands and features disappear, obtaining a flatter spectrum with less definition, hence reaching large sizes where only the KBr spectrum can be observed.
    To obtain well-defined spectra at smaller sizes distribution it is better to use a concentration of 0.5\% (variable up to 1\%).

    For this study, there was an experimental limitation when performing similar measurements with meteorite dust available at the Cosmic Dust Laboratory because the amount of material is limited and there is a high distribution of the initial particle size of each sample. The largest measured sizes for meteorites are less than 100 \textmu m (original size); therefore, for most samples, the spectra in these ranges do not vary much, mainly since the samples of original size contain a large number of small particles. However, it is possible to observe differences in the shape of meteorite spectra of samples with original size (with larger particles) and those grounded for 60 - 75 minutes, as shown in Fig. \ref{fig:los-vientos-abs-size} for LoV123. For grounded samples, the peak of the size distribution is around 3 \textmu m, as shown in Fig. \ref{fig:all-meteorites-sizes}.  

    In this study, we focused on measuring the effect of the grain size distribution on the observed spectra according to the bulk composition of the meteorites. Although crushing bulk meteorites and sieving them is one way to obtain size fractions, the equivalent natural process would be collision/disruption of the parent body and the following grinding sequence producing debris and dust observable in the IR. However, chondrules, CAIs, and matrices existed as separate entities in the circumstellar/protoplanetary disk before being incorporated into larger bodies. 
    Therefore, it will be interesting for future work to obtain spectra from separated samples of these components.

\subsection{Opacities}
\vspace{0.3cm}
    Although this work includes only two meteorite classes, our spectra point toward noticeable heterogeneity regarding mineralogy. Moreover, according to the dichotomy in the composition of the Solar System, carbonaceous meteorites and ordinary chondrites come from different parts of the Solar System \citep{2020SSRv..216...55K}. Although OCs are believed to come from the inner part of the disk, carbonaceous chondrites are believed to come from the outer part of the disk. Even though carbonaceous chondrites were formed in an environment rich in volatile material and ice, CO chondrites typically contain little water, possibly because they were accreted inside the snowline \citep{2024M&PS...59.1170G}. Despite these compositional differences, the variation in the absolute values of the opacities within our samples is around $\sim$ 10 \%. This means that the composition has a greater impact on the shape of the spectra than the absolute value of the opacity. 

    In the case of grain size distribution, although it affects the shape of the spectra, it has an even bigger impact on the total opacity. From Fig. \ref{fig:DSHARP-size} (top panel), it is possible to observe a difference of $\sim$ 20 \% in total opacity between the samples of 3 - 5 \textmu m and 38 - 45 \textmu m for Los Vientos 123. From the bottom panel (olivine), it is possible to observe a decrease in intensity when the dust grains exceed 100 \textmu m.

\begin{figure*}[]
        \centering
                \includegraphics[width=18cm]{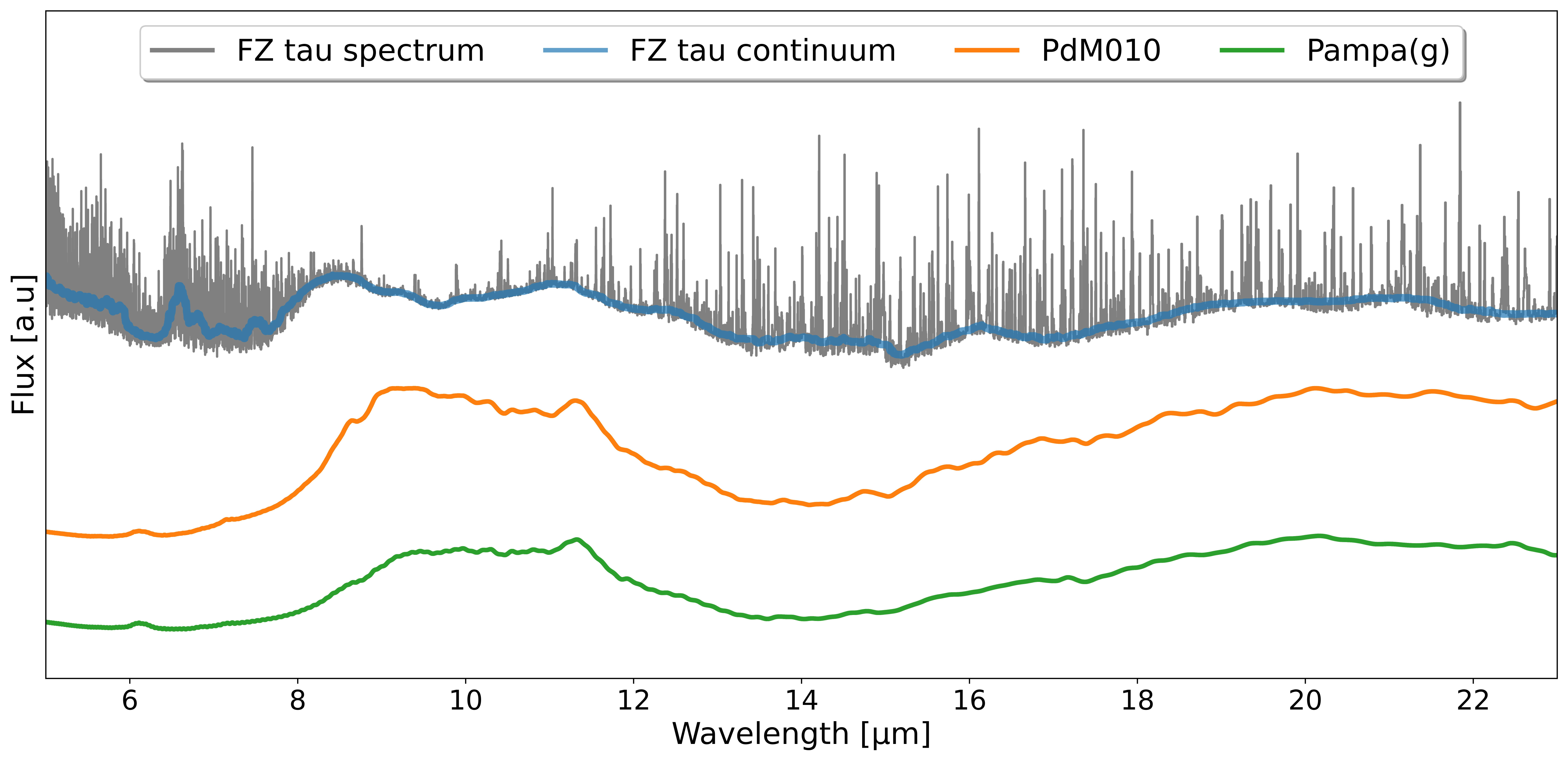}
          \caption{MIRI spectrum of the protoplanetary disk around the young stellar object FZ Tau from \cite{2024ApJ...963..158P}. The spectrum contains both narrow atomic and molecular lines and broad dust features (blue curve). The spectra of two meteorites from this work are shown for comparison.}
   \label{fig:MIRI}
\end{figure*}

\section{Applications}\label{section_5}
\vspace{0.3cm}
    The atmosphere of the Earth becomes very opaque at MIR wavelengths beyond a few microns. Therefore, several generations of space telescopes have been launched in recent decades to fully access the MIR with imaging and spectroscopic instruments. Some major MIR missions include the Infrared Astronomical Satellite (IRAS), the Infrared Space Observatory (ISO), and \emph{Spitzer}. 

    Currently, the JWST is the main MIR space facility. The most relevant instrument for laboratory measurements presented in this work is the Mid Infrared Instrument (MIRI), which has an observing mode for Medium Resolution Spectroscopy (MRS) to observe in wavelengths between 4.9 \textmu m and 28.5 \textmu m. JWST has a primary mirror with an area 50 times larger than \emph{Spitzer}, and MIRI has a higher spectral resolution that goes from R $\sim$ 3300 at 4.9 \textmu m to R $\sim$ 1500 at 28.5 \textmu m. These improvements over previous-generation instruments allow us to make a more detailed comparison of astronomical observations with laboratory measurements. 
    For instance, the MIRI observations of protoplanetary disks are rich in narrow atomic and molecular line data on top of broad dust features. 
    The features of dust are typically compared to those of pure crystalline minerals \citep{2024ApJ...963..158P}, which limits the information that can be extracted from telescopic observations. To address this limitation, measurement of meteorite spectra with varying grain-size distributions can provide better insights into the composition and structure of dust in protoplanetary disks. 
    As shown in Fig. \ref{fig:MIRI}, we present a preliminary and basic comparison between the MIR spectra of Fz Tau and two of our meteorite samples. This comparison serves as an initial example of the potential applications of meteorite spectra in interpreting protoplanetary disk observations. PdM010 exhibits a prominent feature associated with amorphous material, which typically appears in small quantities in meteorites. This particular feature in the meteorite spectra shows a qualitative similarity to a feature observed in Fz Tau, though slightly shifted. While this comparison is preliminary and not exhaustive, it highlights the potential of using meteorite measurements to better interpret spectra of protoplanetary disks.

    Furthermore, effects of grain sizes on the MIR spectra of protoplanetary disks (e.g., \citealt{2006ApJ...639..275K}) are usually treated analytically using Mie theory \citep{1908AnP...330..377M} and assuming spherical grains. Similarly, the porosity of the grains is usually tackled by adopting distributions of hollow spheres \citep{2005A&A...432..909M}.
    More complete and realistic opacity libraries constructed with different types of meteorites with known composition and grain size distributions will allow us to tackle interesting problems related to planet formation, including chemistry, crystallization, dust migration, grain growth, and impacts.

    Current instrumentation also allows us to study dust features as a function of radius in nearby protoplanetary disks. For instance, the MATISSE instrument on the VLTI has a spatial resolution of $\sim$8 mas at 10 \textmu m. At 140 pc, the distance of nearby star-forming regions that correspond to the spatial resolution of $\sim$1 au from the star, well within the Asteroid Belt region in our Solar System, where we expect chondrite-type material \citep{2022A&A...659A.192L}.

    We can make a simple comparison between the spectra of HD 142527, which at r $\leq$ 3 au looks similar to PdM010, while Lutschaunig's Stone is more similar in shape to r $\leq$ 1.5 au. As such, by comparing meteorite MIR data with high-resolution astronomical observations, we can provide a more complete picture of the mineralogy within protoplanetary disks, including at different locations within the disks. We can see how the spectra of the inner disk of HD 142527 show a common type of material that we see in the inner part of the early Solar System. These results could hence prove useful for future models. 

    Instruments such as MATISSE open the window to the study of regions where terrestrial planets form in solar nebula analogs. They can also be used to study the distribution of different materials across the disk, which is very important because the dust in protoplanetary disks is believed to undergo significant radial and vertical mixing. 
    As evidence of this, crystalline olivine, mainly its Mg-rich member forsterite (Mg$_2$SiO$_4$), is detected in the spectra of comets \citep{1999ApJ...517.1034W, 2009P&SS...57.1133K, 2020Icar..33813450O}, which implies that the crystalline forsterite was well distributed in the places where these comets originally formed \citep{leone2023igneous}. However, the formation of crystalline forsterite is thought to occur through thermal annealing of amorphous olivine aggregates, which are one of the basic building blocks of the Solar System. Thermal annealing occurs at temperatures of around $\sim$1000 K \citep{1999M&PS...34..897W, 2002M&PS...37.1579N}, which normally occurs only well within 0.1 au of the Sun. Therefore, the presence of crystalline material in comets requires an efficient radial transport mechanism to transfer this material to the outer regions of the disk, for example, through radial mixing \citep{2002A&A...384.1107B} or winds and outflows \citep{2012ApJ...744..118J}.

    While gas-rich protoplanetary disks are analogous to the solar nebula, circumstellar dust can be observed even when the primordial gas has dissipated. Such is the case with debris disks (Hughes et al. 2018), which are seen around objects that are tens, hundreds, or even thousands of millions of years old (e.g., systems even as old as the Solar System).   
    The study of dust in debris disks was revolutionized by \emph{Spitzer} \citep{2006ApJS..166..351C, 2014ApJS..211...25C, 2007ApJ...658..584L}, making it possible to spectroscopically verify that the dust is presumably produced by the collision of planetesimals. 
    Observations of this processed or "second-generation" dust can also be compared with our spectral library meteorite dust because they might have similar origins and compositions. 
    The resolution of the JWST and the new observations of debris disks allow us to observe these disks in greater detail. An example that demonstrates this, is the detailed spectra of $\beta$ Pictoris obtained by \cite{2024ApJ...964..168W}, showing that it is possible to identify characteristic features around 10 \textmu m of silicates, such as olivine and pyroxene. 

    A different approach to studying circumstellar dust at much later stages of stellar evolution is to observe white dwarfs (WD),  the stellar core remnants that most stars leave behind after the end of the giant phase.   
    A special class of WDs, known as polluted WDs (PWDs), are enriched in heavy elements by asteroids, comets, and/or exo-Kuiper objects that fall on their surface (due to their chaotic orbits and the strong gravitational attraction of the WD), where they are quickly destroyed by the impact and the strong gravity.   
    As a result, spectroscopic observations of PWDs can probe the bulk composition of the objects that have fallen.  
    The optical and ultraviolet spectra of PWDs show indications that the pollutants are mostly made of rocky material with a variable fraction of ice and carbon \citep{2021Eleme..17..241X}. 

    Many PWDs host dusty disks \citep{2016NewAR..71....9F}, and with JWST it is now possible to observe dust features in the MIR spectra of such WD debris disks. Recently, \citet{2024MNRAS.529L..41S} presented the first JWST spectra of a WD debris disk (WD 0145+234),  which is believed to have experienced a major collision in 2018, producing large amounts of dust. 
    Its spectra look very similar to the meteorites presented in this work in the 10 \textmu m region.  Also, WD 0145+234 shows a feature around 7 \textmu m, similar to what we have seen in our CO3 chondrites.  

    Given all of the above, it is clear that MIR spectral libraries from laboratory measurements of pure minerals and meteorites can significantly contribute to our understanding of protoplanetary,  debris, and WD disks.
    Comparison of the observed spectra to these libraries can be used to identify pure minerals and realistic combinations of minerals that can be connected to chemical and physical processes (e.g., thermal annealing, radial mixing, and collisions). In the case of protoplanetary disks, the use of more realistic spectral libraries can also improve the continuum subtraction and allow for more detailed study of overlapping molecular lines. 

    Furthermore, opacity libraries are a key ingredient of radiative transfer codes such as RADMC-3D \citep{2012ascl.soft02015D} and MCFOST \citep{2006A&A...459..797P}, which are usually used to compare models with astronomical observations. 
    In these radiative transfer codes, individual "photon packets" are emitted from the central star and propagate through the disk. The propagation process is managed by scattering, absorption, and thermal re-emission, which are controlled by both the geometry of the circumstellar environment and the optical properties of the medium, mainly of the dust. The amount of absorbed radiation sets the temperature of the dust and gas and defines the amount of radiation that is re-emitted at thermal wavelengths. 

    Opacity is a function of composition, grain size distribution, and wavelength. The opacity libraries used for radiative transfer modeling are usually created from extrapolations of a few laboratory measurements, making strong assumptions about the structure and grain size distribution.
    With more complete libraries, covering a wider parameter space, fewer assumptions and extrapolations are necessary, resulting in more accurate models. The Bruker Vertex 80V spectrometer at the UDP Cosmic Dust Laboratory has cryogenic capabilities that can be used for measurements of cold samples (such as H$_2$O, CO, CO$_2$ ices and iced-covered grains) down to 4 K. It also has a far-IR and a submillimeter detector that allow us to observe at longer wavelengths, overlapping the wavelength ranges observed by the Herschel Space Observatory (55 \textmu m to 670 \textmu m) and ALMA (300 \textmu m to 3 mm). 

    Related to the study of meteorites itself, these databases would be very useful to researchers seeking mineralogical information on individual meteorite samples, adding to the information already present in the meteoritical database.

    Infrared spectra of meteorites could help with some misclassified samples \citep{2018GeCoA.221..406A}. For instance, CMs and COs have been misclassified interchangeably on multiple occasions due to their similar petrographic properties. However, spectrally, CMs tend to differ from COs due to their abundant phyllosilicate content \citep{1980GeCoA..44.1543B, 1993GeCoA..57.3123Z, 2015GeCoA.149..206H}. Another example is ALHA 77003 \citep[CO3;][]{grossman1994meteoritical}, which was originally misclassified as an OC. As our data suggest, COs and OCs show clear spectroscopic differences.

    Finally, we acknowledge the bias inherent in our samples, as discussed in the introduction, where we noted that OCs are overrepresented in meteorite collections due to their higher survivability during atmospheric entry and greater resistance to weathering. Although OCs are the most abundant chondrite class among meteorite finds, they are not representative of the broader compositional diversity of our Solar System, which is dominated by CCs. To address this limitation and provide a more comprehensive understanding of meteoritic materials, we plan to extend our libraries in future work to include a wider diversity of meteorite samples, particularly CCs, as well as to expand the spectral range of our measurements.

\section{Conclusion}\label{section_6}
\vspace{0.3cm}
    We have presented the first results of the UDP Cosmic Dust Laboratory, which are aimed at contributing to the development of libraries of absorption spectra and opacities of dust for astronomical applications. Based on our initial results, we derive the following conclusions: 
 
\begin{enumerate}

    \item The measurement of absorbance spectra and absolute opacities for meteorite dust with a known composition and particle size distribution requires detailed laboratory protocols for sample preparation (e.g., the preparation of dry KBr pellets with the right concentration of dust) and characterization (e.g., microphotography for the determination of grain sizes). \\

    \item The similarity of our absorbance spectra and opacities to previous studies validates the general procedure performed at the UDP Cosmic Dust Laboratory. \\

    \item Characterization of the primary mineralogy of meteorites is possible through their absorbance spectra, where approximate compositions of different mineral phases can be inferred, such as olivine, pyroxene, and plagioclase. \\

    \item For CO3 chondrites, even though the amount of carbonaceous material is low, the absorption bands related to carbon are very well defined and easy to identify. We aim to perform a more exhaustive study of MIR spectra in the fluid inclusions of these samples. \\

    \item The OH and H$_2$O bands are also easily identifiable. In a more detailed study of this aspect, the degrees of weathering and alteration of the samples, together with the secondary and hydrated minerals, could be obtained from the MIR spectra in conjunction with Raman spectroscopy. This will be considered for future analysis.  \\

    \item From the MIR spectra, it is possible to identify meteorites with amorphous material (glass). This aspect could enable studies of the pristine glass material or shock metamorphism during disk evolution. \\

    \item The effects of particle size on the absorbance spectra and the opacities are very significant, highlighting the importance of performing measurements with a wide range of grain sizes. \\

    \item The MIR spectral and opacity libraries have direct applications for astronomical observations, particularly for some of the most advanced instruments on the ground (VLTI-MATISSE) and in space (JWST-MIRI). Such applications can have a promising impact on the modeling of observable spectra of protoplanetary, debris, and WD disks, as we briefly discussed in Sect. \ref{section_5}.   \\

    \item Tantalizing future directions for new libraries include obtaining samples of meteorites from different classes, extending the measured range to longer wavelengths, measuring samples at different temperatures, and deriving the measurements of astrophysical relevant ices (e.g., H$_2$O, CO, and CO$_2$) and ice covered grains. \\

\end{enumerate}

We expect the data presented in this study will serve as a valuable resource for the astronomy community, contributing to more accurate simulations and a deeper understanding of planetary formation. Looking ahead, we plan to expand meteorite measurements and analyses in the UDP Cosmic Dust Laboratory, further advancing our knowledge of extraterrestrial materials.

\vspace{0.5cm}
\section*{Data Availability}
The first MIR (2-23 \textmu m) libraries from the laboratory are publicly available on \href{https://zenodo.org/records/14919561}{Zenodo} (\href{https://doi.org/10.5281/zenodo.14919560}{https://doi.org/10.5281/zenodo.14919560}). Additional data and resources can be accessed at the \href{https://astronomia.udp.cl/en/collaboration/udp-cosmic-dust-laboratory/}{Cosmic Dust Laboratory} webpage (\href{https://astronomia.udp.cl/en/collaboration/udp-cosmic-dust-laboratory/}{https://astronomia.udp.cl/en/collaboration/udp-cosmic-dust-laboratory/}).

\vspace{0.5cm}
\begin{acknowledgements}
    We express our gratitude to the referee for his thoughtful and constructive comments. His valuable insights and suggestions have greatly contributed to improving the quality and clarity of the manuscripts. We greatly appreciate the time and effort that he dedicated to the review process.
    
    The UDP Cosmic Dust Laboratory was primarily funded by the QUIMAL 150011 grant from ANID (Agencia Nacional de Investigacion y Desarrollo) of Chile.
    
    G.A.B.F was supported by the ANID FONDECYT Grant \# 1211656.  L.A.C  acknowledges support from the Millennium Nucleus on Young Exoplanets and their Moons (YEMS), ANID - Center Code NCN2021$\_$080, and the FONDECYT grants \# and 1211656 and 1241056.

\end{acknowledgements}


\end{document}